\documentclass[traditabstract]{aa}

\usepackage{newtxtext,newtxmath}
\usepackage{natbib}

\usepackage[T1]{fontenc}
\usepackage{ae,aecompl}

\usepackage{graphicx}	
\usepackage{amsmath}	
\usepackage{amssymb}	
\usepackage{ulem}
\usepackage{url}

\begin{document}

\title{Signatures from a merging galaxy cluster and its AGN population: LOFAR observations of Abell 1682.}   \titlerunning{LOFAR observations of A1682}

\author{A. O. Clarke\inst{1}, A.~M.~M.~Scaife\inst{1}, T. Shimwell\inst{3,2}, R. J. van Weeren\inst{2}, A. Bonafede \inst{4}, G. Heald\inst{5,3,6}, G. Brunetti\inst{4}, T. M. Cantwell\inst{1}, F. de Gasperin\inst{7}, M. Br\"uggen\inst{7}, A. Botteon\inst{4}, M. Hoeft\inst{10}, C. Horellou\inst{8}, R. Cassano\inst{4}, J. J. Harwood\inst{9}, H.J.A. R\"ottgering\inst{2}} 
\authorrunning{Clarke et al.}
\offprints{A. O. Clarke, \email{alex.clarke-3@manchester.ac.uk}}

\institute{Jodrell Bank Centre for Astrophysics, School of Physics \& Astronomy, University of Manchester, UK, M13 9PL \and 
Leiden Observatory, Leiden University, PO Box 9513, NL-2300 RA Leiden, the Netherlands \and 
ASTRON, the Netherlands Institute for Radio Astronomy, Postbus 2, 7990 AA, Dwingeloo, The Netherlands \and 
Dipartimento di Fisica e Astronomia, Università di Bologna, via P. Gobetti 93/2, 40129 Bologna, Italy INAF-IRA, via P. Gobetti 101, 40129 Bologna, Italy \and 
CSIRO Astronomy and Space Science, 26 Dick Perry Avenue, Kensington WA 6151, Australia \and 
Kapteyn Astronomical Institute, University of Groningen, PO Box 800, 9700 AV, Groningen, The Netherlands \and 
Hamburger Sternwarte, Universitat Hamburg, Gojenbergsweg 112, 21029, Hamburg, Germany \and 
Chalmers University of Technology, Dept of Space, Earth and Environment, Onsala Space Observatory, SE-439 92 Onsala, Sweden \and 
Centre for Astrophysics Research, School of Physics, Astronomy and Mathematics, University of Hertfordshire, College Lane, Hatfield AL10 9AB, UK \and 
Th\"uringer Landessternwarte, Sternwarte 5, 07778 Tautenburg, Germany 
}

\abstract{
We present LOFAR data from 110--180~MHz of the merging galaxy cluster Abell 1682, alongside archival optical, radio and X-ray data. Our 6 arc-second resolution images at low frequencies reveal new structures associated with numerous radio galaxies in the cluster. At 20 arc-second resolution we see diffuse emission throughout the cluster over hundreds of kpc, indicating particle acceleration mechanisms are in play as a result of the cluster merger event and powerful active galactic nuclei. We show that a significant part of the cluster emission is from an old radio galaxy with very steep spectrum emission (having a spectral index of $\alpha < -2.5$). Furthermore we identify a new region of diffuse steep spectrum emission ($\alpha < -1.1$) as a candidate for a radio halo which is co-spatial with the centre of the cluster merger. We suggest its origin as a population of old and mildly relativistic electrons left over from radio galaxies throughout the cluster which have been re-accelerated to higher energies by shocks and turbulence induced by the cluster merger event. We also note the discovery of six new giant radio galaxies in the vicinity of Abell 1682.}



\maketitle

\section{Introduction}

Clusters of galaxies are the largest gravitationally bound structures in the Universe and arise primarily at filament junctions in the cosmic web. Under the influence of strong gravitational fields at these junctions, it becomes increasingly likely that clusters will collide or merge, which can release $10^{64}$ erg of gravitational potential energy into the intra cluster medium. This energy heats up the gas interwoven between the hundreds of galaxies and produces shocks and turbulence throughout the intra-cluster medium \citep[ICM;][]{brunetti2014}.

Owing to the presence of large-scale micro-Gauss-level magnetic fields throughout the cluster, the electrons in the ICM can produce synchrotron radiation at radio wavelengths (see \cite{weerenreview2019} for a recent review). Different acceleration mechanisms produce quite different observational features, namely polarised elongated structures called relics, and diffuse low surface brightness unpolarised structures called halos. A combination of diffusive shock acceleration \citep{blandford1987} and shock drift acceleration \citep{ensslin1998,guo2014a,guo2014b} can explain the observed energetics of many relics, but other examples \citep{weeren2016toothbrush,botteon2016,eckert2016} require additional constraints where the seed electrons are already mildly relativistic instead of being accelerated from the thermal pool. A prominent example of shock acceleration is shown by \cite{weeren2017} who present a relic being connected and illuminated by the jet of a radio galaxy. 

In the favoured model for radio halo formation, a seed population of old relativistic electrons or secondary particles are re-accelerated to ultra relativistic energies (GeV) by turbulence induced by a cluster merger event \citep[e.g.][]{brunetti2001,petrosian2001,BrunettiLazarian2007,BrunettiLazarian2011,brunetti2014,pinzke2015,pinzke2017}. This process also gives rise to a relatively new and elusive population of Ultra Steep Spectrum Radio Halos \citep[USSRH;][]{brunetti2008nat, cassano2006} that show up at very low frequencies. \noindent Furthermore, throughout galaxy clusters there are often many radio galaxies, which can produce very bright synchrotron emission from relativistic jets that extend throughout the cluster. These radio galaxies are candidates for providing a seed population of mildly relativistic electrons, and there is evidence that shocks from cluster merger events have interacted with the tails of old radio galaxies \citep{ensslin2001, weeren2017, gasperinGREET2017}.

The LOw Frequency ARray \citep[LOFAR;][]{LOFAR} is a telescope based in the Netherlands, and is an effective instrument for observing synchrotron radiation in galaxy clusters, as shown by recent observations from \cite{weeren2016toothbrush,shimwell2016,hoang2017,wilber2017,Botteon2018}. The results of this paper are based on observations performed with LOFAR's high-band antennas (HBA) at 110--180 MHz. With an excellent \textit{uv} coverage, and operation at these low frequencies, LOFAR has the novel capability of detecting and resolving at arc-second resolution the synchrotron emission that we expect to see in the ICM of merging clusters.

This paper is organised as follows. In Section \ref{section:a1682} we introduce Abell 1682, and summarise previous studies in Section \ref{section:previousstudies}. In Section \ref{section:data} we present new LOFAR observations, archival GMRT \citep[Giant Meterwavelength Radio Telescope;][]{GMRT}  observations, and analysis of the spectral index. In Section \ref{section:analysis} we analyse and discuss the observational data, and we draw conclusions in Section \ref{section:conclusions}. Throughout this paper we use the J2000 coordinate system, and adopt the spectral index convention that $S\propto\nu^\alpha$, where $S$ is the flux density, $\nu$ is the frequency, and $\alpha$ is the spectral index. We assume a flat Universe with \mbox{${\Omega_\mathrm{m}} = 0.286$}, \mbox{${\Omega_\Lambda} = 0.714$}, and \mbox{${\textrm{H}}_0 = 67.8\,\textrm{km} \,\textrm{s}^{-1} \textrm{Mpc}^{-1}$} \citep{plank2016}, unless explicitly stated otherwise. 

\subsection{Abell 1682}\label{section:a1682}

\begin{figure*}
\includegraphics[width=0.5\hsize]{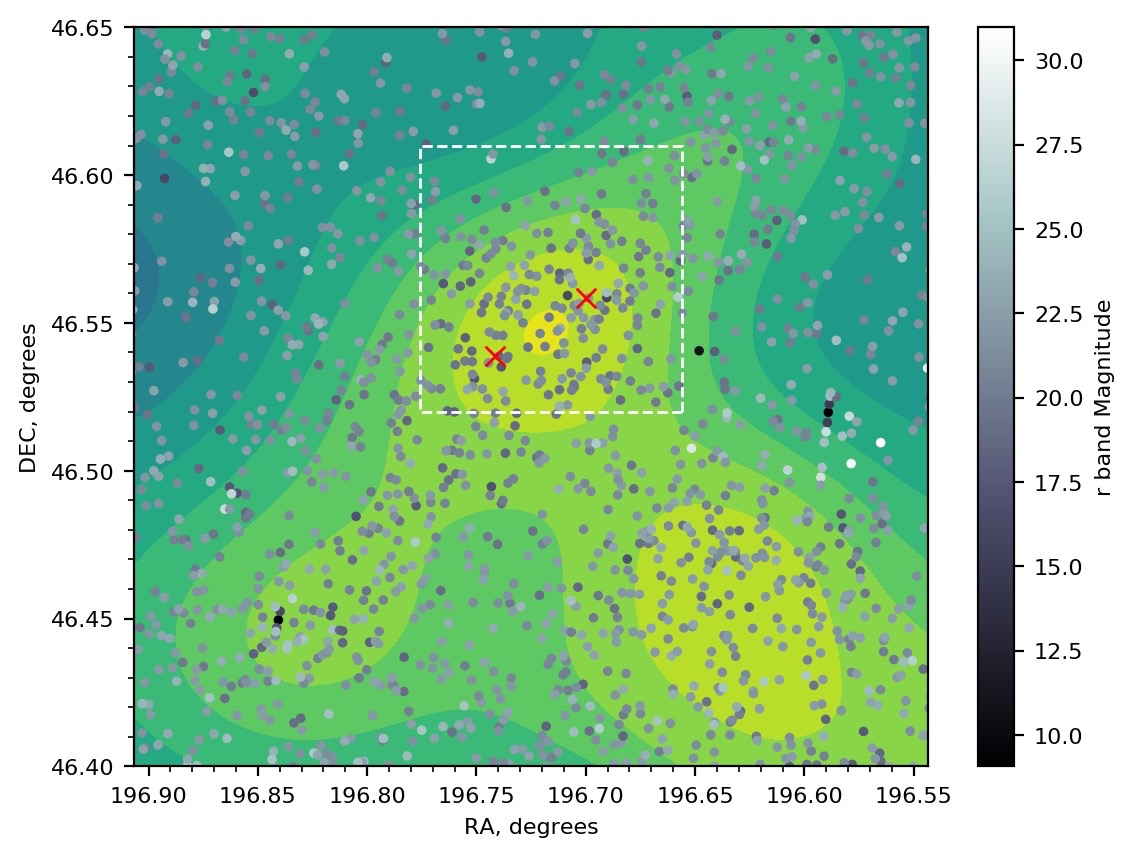}
\includegraphics[width=0.46\hsize]{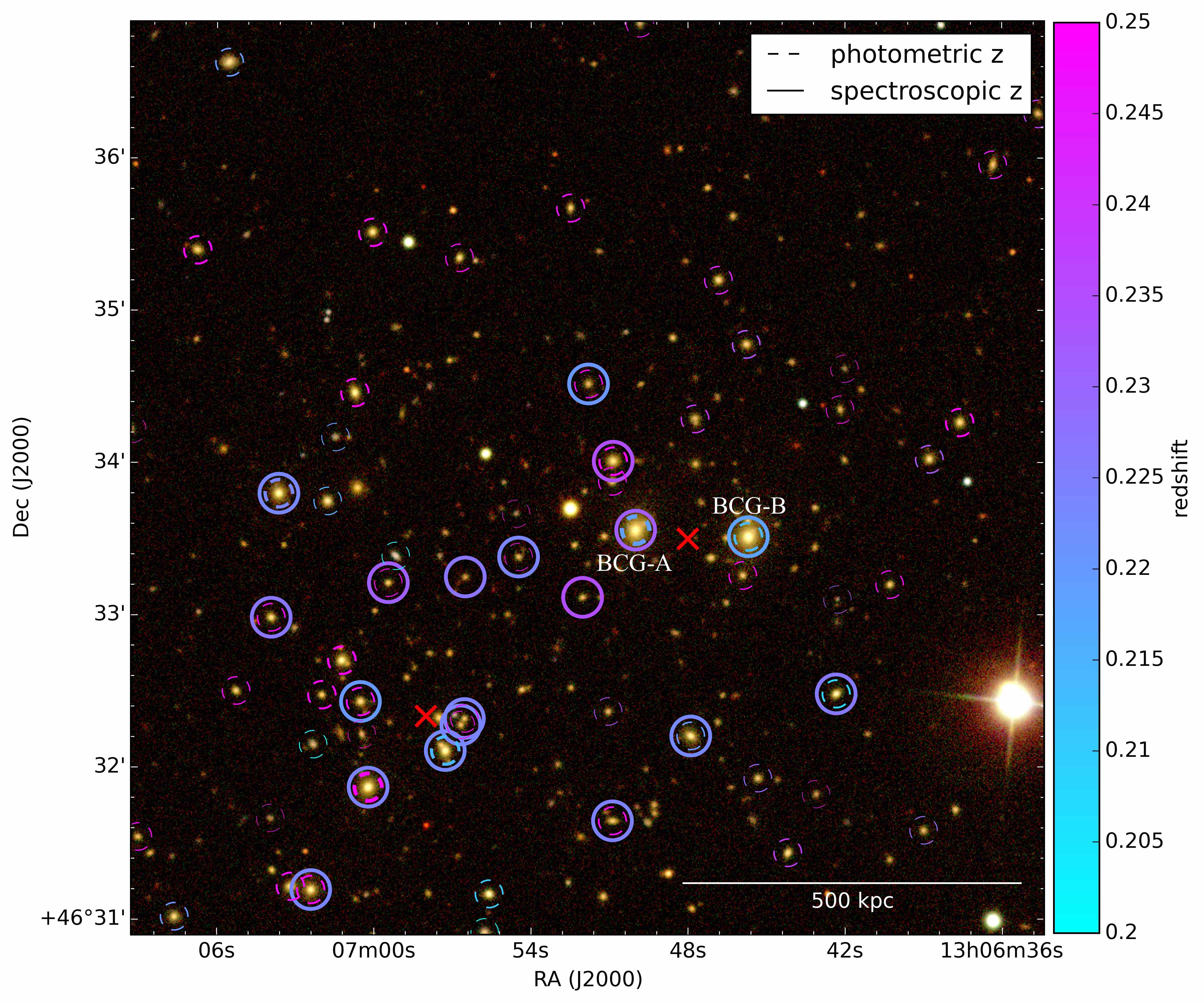}
\caption{Left: Dots showing all galaxies identified in the region from SDSS, with the corresponding galaxy density shown by filled contours. Red crosses denote the centre of mass of the sub-clusters identified by \protect \cite{dahle2002}. Right: SDSS composite image from bands \textit{g}, \textit{r}, \textit{i} within the area shown by the white square in the left plot. The two brightest galaxies are labelled, where BCG-A (13:06:49.99, +46:33:33.35) has a redshift of 0.232, and and BCG-B (13:06:45.69, +46:33:30.74) has a redshift of 0.218. The width of the dashed circles around galaxies with photometric redshifts is proportional to the error on the measurement, where they start at 4\% and get thinner as the error increases. The maximum photometric error is 50\%.}
\label{figure:galdistribution}
\end{figure*}

\begin{figure}
\includegraphics[width=\hsize]{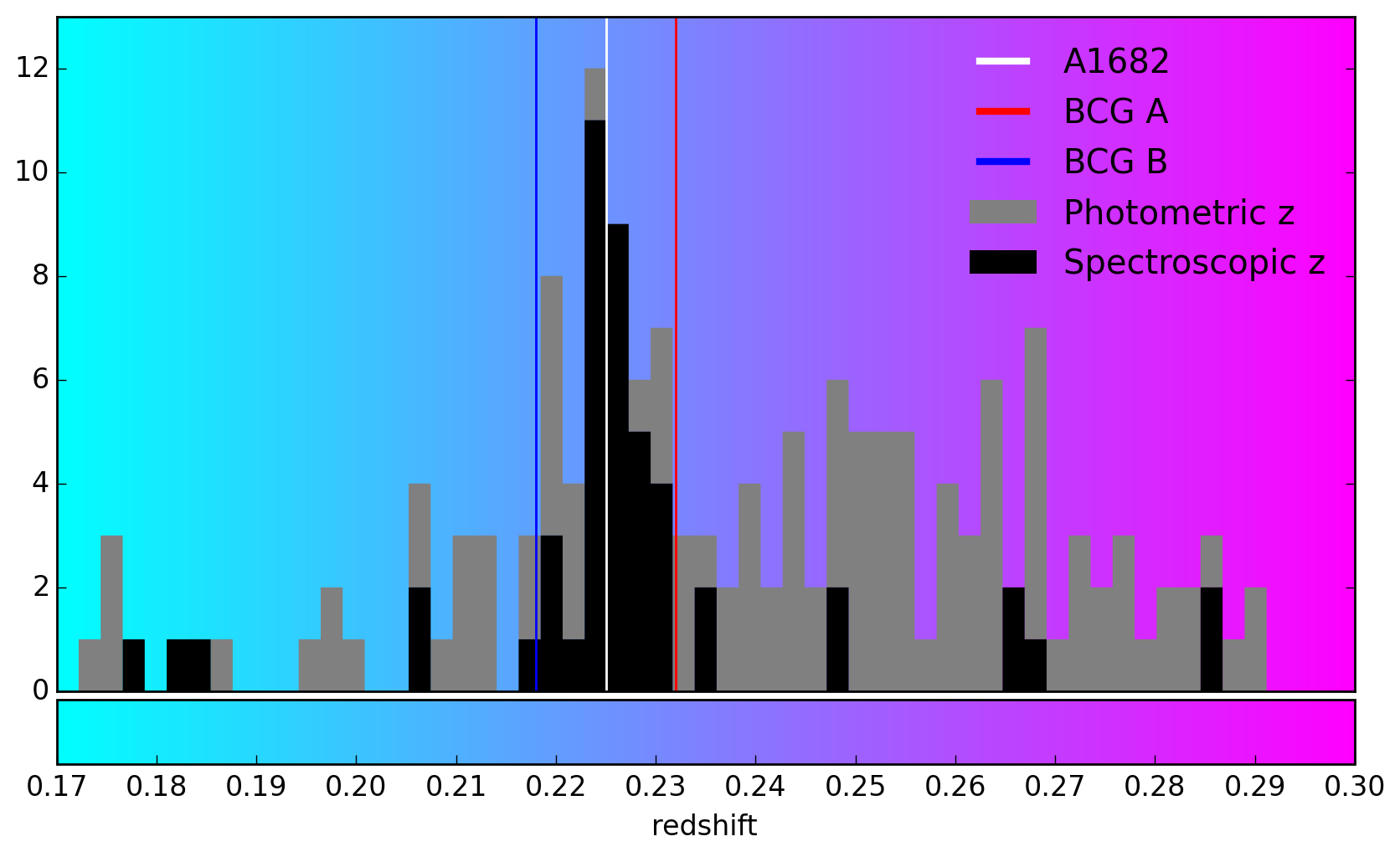}
\caption{A stacked histogram of galaxies with photometric and spectroscopic redshifts within a 6\arcmin radius around A1682. The redshift of A1682 is calculated as the average of BCG-A and BCG-B, which aligns well with the peak of the histogram.}
\label{figure:hist}
\end{figure}

A1682 is a massive merging galaxy cluster, with an X-ray luminosity $L_X~\mathrm{[0.1-2.4 \, keV]} = 11.26 \times 10^{44}~\mathrm{erg\, s^{-1}}$ \citep[assuming \mbox{${\textrm{H}}_0 = 50\,\textrm{km} \,\textrm{s}^{-1} \textrm{Mpc}^{-1}$};][]{ebeling1998}. It has a total mass of the order of $10^{15}~\mathrm{M}_{\odot}$ and is shown to have a bi-modal mass distribution indicating it is a pre- or post-merger system \citep{dahle2002}. The mass inferred from the Sunyaev-Zeldovich effect is $\mathrm{M}_{500}=6.20 \substack{+0.45 \\ -0.46} \times 10^{14}~\mathrm{M}_{\odot}$ and $\mathrm{M}_{500}=5.70 \pm0.35 \times 10^{14}~\mathrm{M}_{\odot}$ after correction for diffuse galactic plane emission \citep{plankmass2013, plankmass2015}. The left panel in Figure \ref{figure:galdistribution} shows surrounding galaxies identified in the Sloan Digital Sky Survey \cite[SDSS;][]{SDSS} Data Release 13 \cite[DR13;][]{SDSSDR132017} along with a density plot. This agrees with \cite{dahle2002} and shows a significant galaxy density enhancement approximately 1~Mpc to the south-east of A1682. 

The cluster has two dominant elliptical brightest cluster galaxies (BCGs), marked in the right panel of Figure \ref{figure:galdistribution}. BCG-A (13:06:49.99, +46:33:33.35) is at a redshift of 0.232, has a velocity dispersion of $298 \pm 9$ km/s, and an estimated mass of $\textrm{Log}\, (\textrm{M}_\textrm{gal}/\textrm{M}_{\odot}) = 11.93$. BCG-B (13:06:45.69, +46:33:30.74) is at a redshift of 0.218, has a velocity dispersion of $332 \pm 13$ km/s, and an estimated mass of $\textrm{Log}\, (\textrm{M}_\textrm{gal}/\textrm{M}_{\odot}) = 11.81$. These spectroscopic redshifts and masses are taken from SDSS DR13 \citep{SDSSDR132017}, where the redshift errors are $1 \times 10^{-5}$. Figure \ref{figure:hist} shows a histogram of all spectroscopically and photometrically identified galaxies within a $6\arcmin$ radius of the cluster centre. We take the redshift of this cluster as the average of the two BCGs, to be $\textrm{z} = 0.225$ (Figure \ref{figure:hist}), which matches well with the peak of the redshift histogram in Figure \ref{figure:hist}. Older spectroscopic measurements of these galaxies show a very similar average cluster redshift of $\textrm{z} = 0.226$ \citep{allen1992}.

Galaxies identified with photometric redshifts have a mean error of 18\%, where the lowest is 4\%, and the highest 50\% \citep{SDSS}. Despite the 4\% error on the photometric redshift for BCG-A, it is significantly different to the spectroscopic redshift (see the right panel of Figure \ref{figure:galdistribution}). Furthermore, as shown in Figure \ref{figure:hist}, overall the photometric redshift measurements are biased towards higher redshifts. Many of them have significant flags in the training set coverage and fitted parameters \citep{beck2016}, making their error estimate unreliable. Despite these caveats, they do still allow a basic identification of the brightest cluster members in the region. Given these problems with the photometric redshifts of cluster members, in figures throughout this paper we only highlight cluster member galaxies with photometric redshifts in the range 0.2 - 0.25. However, many of the galaxies in the vicinity of Abell 1682 are resolved, appearing as likely cluster members even with photometric redshifts much larger than the average cluster redshift (up to z=0.3, as seen in Figure \ref{figure:hist}).

\subsection{Previous studies}\label{section:previousstudies}
Previous radio studies \citep{venturi2008,venturi2011,venturi2013,macario2013} show that A1682 hosts complex cluster-wide radio emission without a consensus on its origin. \cite{venturi2008} present X-ray data, indicating that there are two sub-clusters which have passed through each other in a north-west to south-east direction (a post-merger scenario). The merger axis appears to be in the plane of the sky, however there is no evidence on what the 3-dimensional velocity and momentum distribution is between the two clusters. \cite{venturi2008} also show the first deep high resolution radio observations at 610~MHz using the GMRT, revealing extended emission (100-500 kpc) resembling radio jets or radio relics, but without a confident assessment of their origin.

Follow-up observations at 240~MHz with the GMRT \citep{venturi2013} show structures very similar to those seen at 610~MHz. They also show higher resolution (1.1\arcsec) VLA data at 1.4~GHz of BCG-B showing clear evidence of radio jets (less than 50 pc in size) emanating from an AGN in the centre of this galaxy (see their Figure 7). In their VLA data the radio lobes are brightest on the north-east side, becoming fainter towards the south-west, potentially suggesting that this radio galaxy (BCG-B) is moving north-east. However, the radio lobes could simply be expanding into a less dense region of the cluster towards the south-west whilst the host galaxy does not move. Whilst radio jets can significantly change direction in a dense cluster environment, it is conceivable that the two 500 kpc radio features connected to BCG-B are jets and lobes originating from this AGN. However, their spectral index map from 240~MHz to 610~MHz conflicts with this argument, as the N-W ridge shows a spectral index gradient perpendicular to the major axis of the structure from $\alpha \approx -1$ to $\alpha \approx -2.3$. This is indicative of a radio relic, where the spectrum is flatter at the shock front ($\alpha \approx -0.6$), and then steepens ($\alpha < -1$) in its wake \citep[e.g.,][]{stroe2014}. The orientation of this structure is perpendicular to the merger axis identified by the X-ray and galaxy distribution data, however, a complex merger scenario and line-of-sight projection affects do not rule out the possibility that this is a relic.

The data at 240~MHz also reveals new emission not seen at higher frequencies, with the detection of an area of diffuse radio emission, which becomes significant at lower resolution. This component is also seen in the Very Large Array Sky Survey \citep[VLSSr;][]{VLSSr2014} at 74~MHz, where \cite{venturi2013} calculate a spectral index between these two frequencies integrating over all the observed emission, with a value of $\alpha = -2.09 \pm 0.15$. However, we caution against using this value due to the different spatial scales that the GMRT and VLA probe given their configurations. Analysis by \cite{venturi2011} attempts to subtract the high resolution data from both the 240 and 610~MHz data, showing significant residuals over the whole cluster, particularly coincident with the diffuse component. The most significant detection of the diffuse component is from observations at 150~MHz with the GMRT \citep{macario2013}. Despite the low sensitivity of this observation (with an RMS noise of 1.7 mJy/beam), the diffuse component is clearly detected, with an integrated flux of $98 \pm 20$ mJy. \cite{macario2013} present a spectral index integrated over the whole emission from 150~MHz to 240~MHz of  $\alpha = -1.7 \pm 0.1$, which is flatter than the value calculated from 240 to 610~MHz, and they suggest that this is due to the lower surface brightness sensitivity of observations at 150~MHz.

\section{Data}\label{section:data and calibration}\label{section:data}

\subsection{LOFAR Calibration and Imaging}
The LOFAR HBA observation presented here used 23 of the core Dutch stations (each comprising two sub-stations) and 13 remote Dutch stations. The central frequency was 150.0~MHz, with a total bandwidth of 70.3~MHz from 111.5~MHz to 181.8~MHz. The primary flux calibrator (3C295) was observed for 5 minutes and the target (13:06:49.70 +46:32:59.00) was observed with a total integration time of 9.67 hours. The separation between the calibrator and target sources is 11.9 degrees. The observation started on 19 April 2013 at 19:06:02.50, and finished on 20 April 2013 at 04:46:00.80 (UTC), however all data after 03:00:00 (CET) were flagged due to very poor ionospheric conditions. In total we used 8 hours of data in the calibration and imaging. We used the flux scale defined by \cite{scaifeheald2012} for 3C295, where the error on the flux scale for 3C295 is 4\% for LOFAR HBA frequencies. However as discussed in \cite{shimwell2019} we take a conservative estimate of the flux scale uncertainty as 20\% given the inadequate knowledge of the LOFAR beam during calibration, and note that sources crossmatched with the Tata GMRT Sky Survey \citep[TGSS-ADR1;][]{intema2017} are consistent within this error. 

We follow the calibration and imaging scheme described in detail in \cite{shimwell2019} which accounts for direction dependent effects and wide field spectral deconvolution. In summary, this begins with a direction-independent calibration (described in detail by \citealt{reinout2016} and \citealt{williams2016}) and makes use of the LOFAR Default Preprocessing Pipeline \citep[DPPP:][]{Diepen2018} for averaging and calibration and BlackBoard Selfcal \citep[BBS;][]{Pandey2009} for calibration. For the direction dependent calibration, KillMS \citep[kMS:][]{Tasse2014, Smirnov2015} is used to calculate the Jones matrices and DDFacet \citep{Tasse2018} is used to apply these during the imaging. The main software packages and the pipeline are publicly available and documented\footnote{see \url{https://github.com/saopicc} for kMS and DDFacet, and
\url{https://github.com/mhardcastle/ddf-pipeline} for the associated LoTSS-DR1 pipeline.}.

Figure \ref{figure:lofar} shows the high resolution LOFAR HBA image. This used a Briggs \citep{Briggs} robust parameter of -0.5 to achieve a resolution of $6.0\arcsec \times 6.0\arcsec$, with an RMS noise of $90\,\mathrm{\mu Jy/beam}$, a \textit{uv}-minimum of $0.1\,\mathrm{k}\lambda$ and no \textit{uv}-maximum limit. There are no significant direction dependent errors visible and the RMS noise is uniform around the source, measured at $90\,\mathrm{\mu Jy/beam}$. The brightest pixel in the image is 0.29 Jy/beam (the peak of the colour-scale). There are small spurious areas with negative wells at -5 times the RMS noise around the brightest part of the cluster.

Figure \ref{figure:lofarlowres} shows the low resolution image. This used a Briggs robust parameter of -0.25, and a \textit{uv}-range of $0.1-25.75\,\mathrm{k}\lambda$, to achieve a resolution of $20\arcsec \times 20\arcsec$. The RMS noise is $130\,\mathrm{\mu Jy/beam}$, and the brightest pixel has a value of 0.74 Jy/beam. As in the high resolution image, there are small spurious areas with negative wells at -5 times the RMS noise around the brightest part of the cluster. At this lower resolution, extended emission is recovered more readily, and the patchy areas of emission seen at high resolution are robustly detected as areas of low-surface brightness diffuse emission. This is most prominent in the region between the clusters two centres of mass, and in regions towards the extremities of the extended structures. 

Overall, we observe all the features identified in previous studies, plus some additional features. The diffuse component is now resolved, and distinguished from diffuse emission in the region between the centre of mass of each of the sub-clusters. In section \ref{section:analysis} we discuss the origin of these components.

\begin{figure}
\includegraphics[width=\hsize]{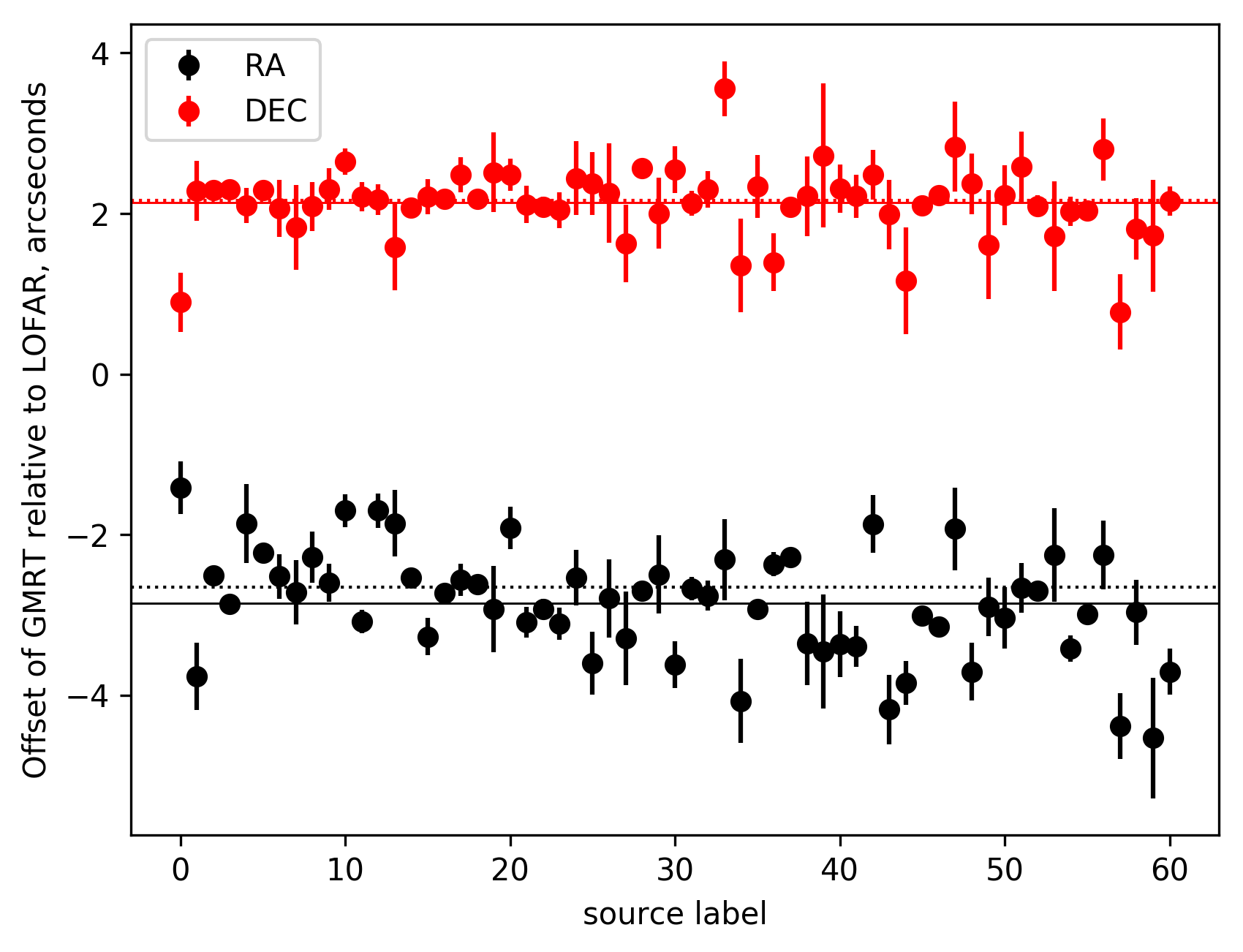}
\caption{The offset in RA and DEC of 61 sources crossmatched between the GMRT and LOFAR. The dashed line is the mean offset calculated from the original 92 sources, and the solid lines show the mean offset calculated after disregarding all sources more than 5 times their position error away from this mean in either RA or DEC. The final offset in RA and DEC was calculated to be 2.86 and 2.14 arc-seconds respectively, and this correction was applied to the GMRT image.}
\label{figure:astrometry}
\end{figure}

\subsection{Spectral index maps with the GMRT}
We re-analysed archival GMRT data at 610~MHz using the Source Peeling and Atmospheric Modelling software \cite[SPAM;][]{intema2014}. Both the GMRT and LOFAR data sets were imaged using a \textit{uv}-minimum of $0.1~\textrm{k}\lambda$ at the same resolution of $6\arcsec \times 6 \arcsec$. As described in \cite{shimwell2019} the astrometry of our LOFAR images has been matched to optical sources in PanSTARRS \citep{panstarrs2016}, however our GMRT image was mis-aligned, and so we applied an astrometry correction to our GMRT image to align it with our LOFAR image. To do this we first masked A1682 and used PyBDSF \citep{pybdsf2015} to detect sources above 5 times the RMS noise in both the GMRT and LOFAR images. We then crossmatched all sources detected by the GMRT with LOFAR using 6 arc-second matching radius. From the 92 sources detected by both the GMRT and LOFAR around A1682, we calculating a mean offset in RA and DEC of -2.65 and 2.16 arc-seconds respectively (dashed black and red lines in Figure \ref{figure:astrometry}). We then disregarded all source that were more than 5 times their position error away from this mean (in either RA or DEC), leaving the 61 sources shown in Figure \ref{figure:astrometry}, and repeated the calculation of the mean offset in RA and DEC giving 2.86 and 2.14 arc-seconds respectively (the solid black and red lines in Figure  \ref{figure:astrometry}). This procedure ensured we only used reliably matched sources to correct the astrometry, and the final correction was applied to the GMRT image.

We used the Broadband Radio Astronomy ToolS \citep[BRATS;][]{harwood2013,harwood2015} to calculate the spectral index value for each pixel and corresponding error map, as shown in Figure \ref{figure:spectralindexmap}. Only data above 5 times the RMS noise in both data sets was used. We note that applying the \textit{uv}-minimum of $0.1~\textrm{k}\lambda$ during imaging was done to ensure both telescopes were sampling the same spatial scales, since LOFAR has significantly better \textit{uv} coverage below this value. However, when doing this we only noticed a change in the spectral index value of around 1\%, indicating that the spatial scales involved in this source were not large enough to greatly affect the spectral index calculations.

\begin{figure*}
\includegraphics[width=0.96\hsize]{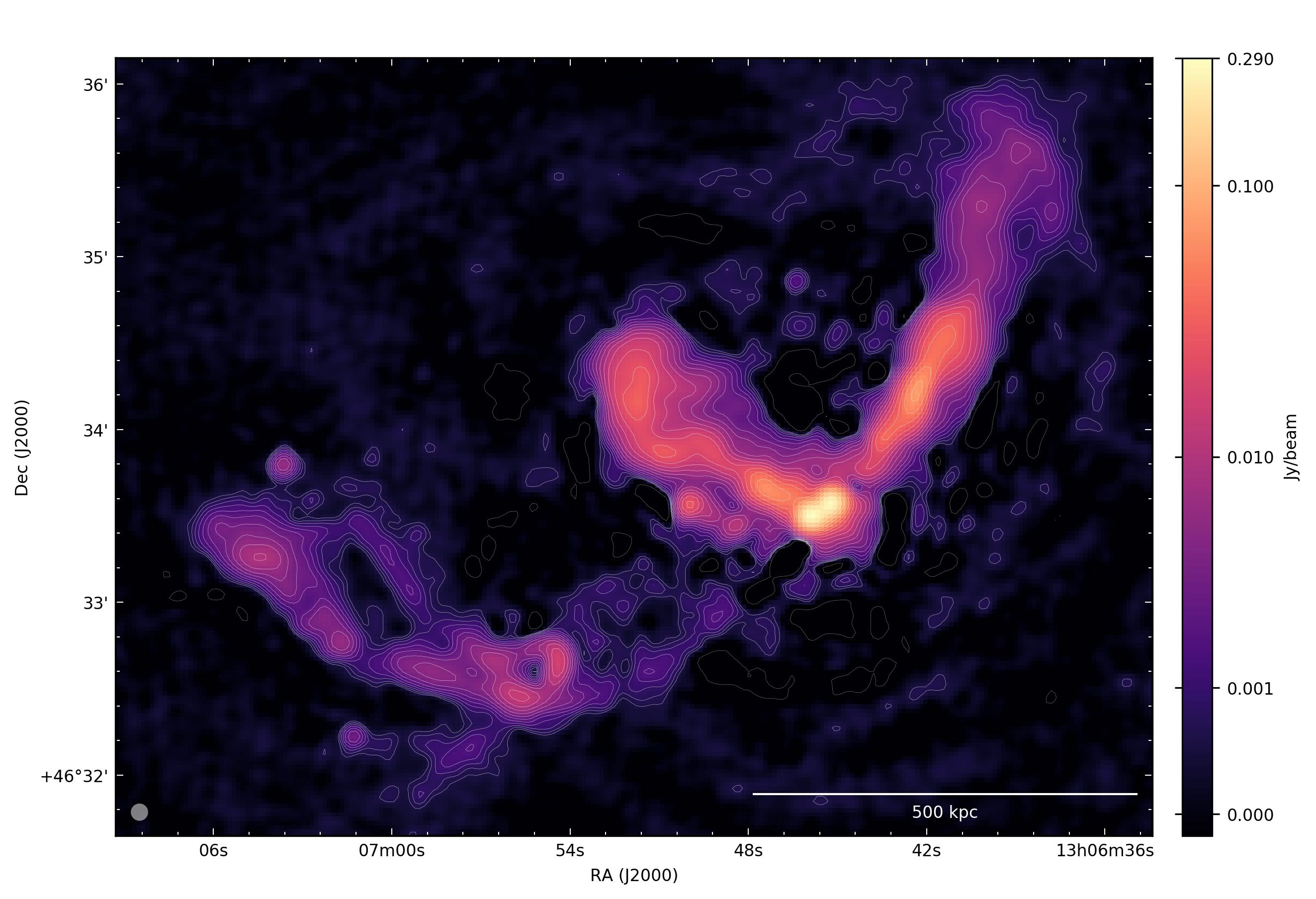}
\includegraphics[width=0.96\hsize]{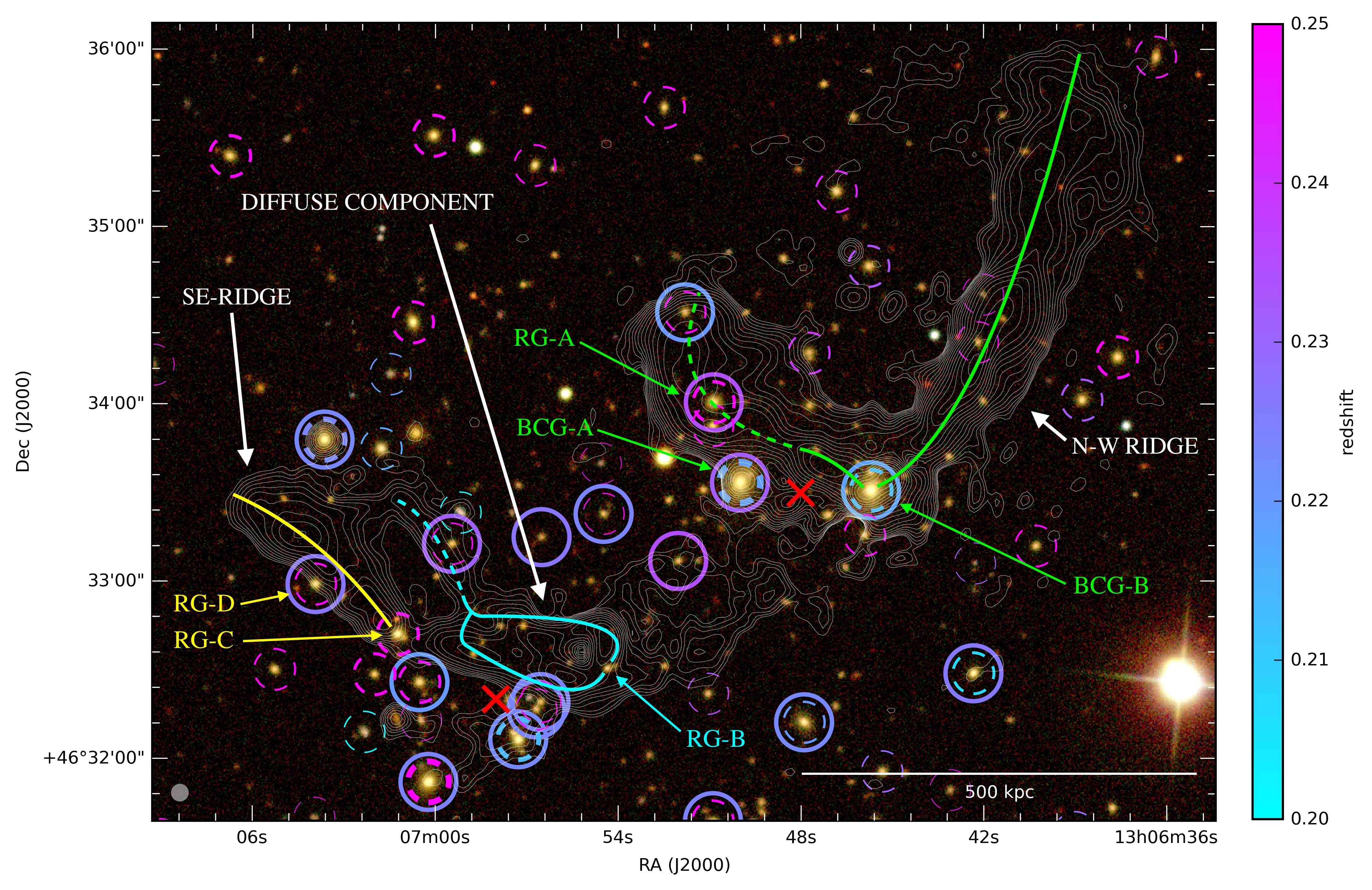}
\caption{Top: Colour-scale image of A1682 on a log stretch from $-90~\mu \textrm{Jy/beam}$ to 0.29~Jy/beam (the brightest pixel is 0.29~Jy/beam). The RMS noise is $90~\mu \textrm{Jy/beam}$. Dashed contours are overlaid at -5 times the RMS noise. Solid contours start at 5 times the RMS noise and increase by factor 1.4. The resolution is $6\arcsec \times 6\arcsec$. Bottom: SDSS background composite image from bands \textit{g}, \textit{r} and \textit{i}, with the same contours as above. Redshift coloured markers are the same as Figure \ref{figure:galdistribution}. Radio galaxy candidates (RG-A to RG-D) are marked, along with lines showing the possible orientation of their radio jets/tails. Red crosses mark the centres of mass of the two sub-clusters identified by \protect \cite{dahle2002}.}
\label{figure:lofar}
\end{figure*}

\begin{figure*}
\includegraphics[width=\hsize]{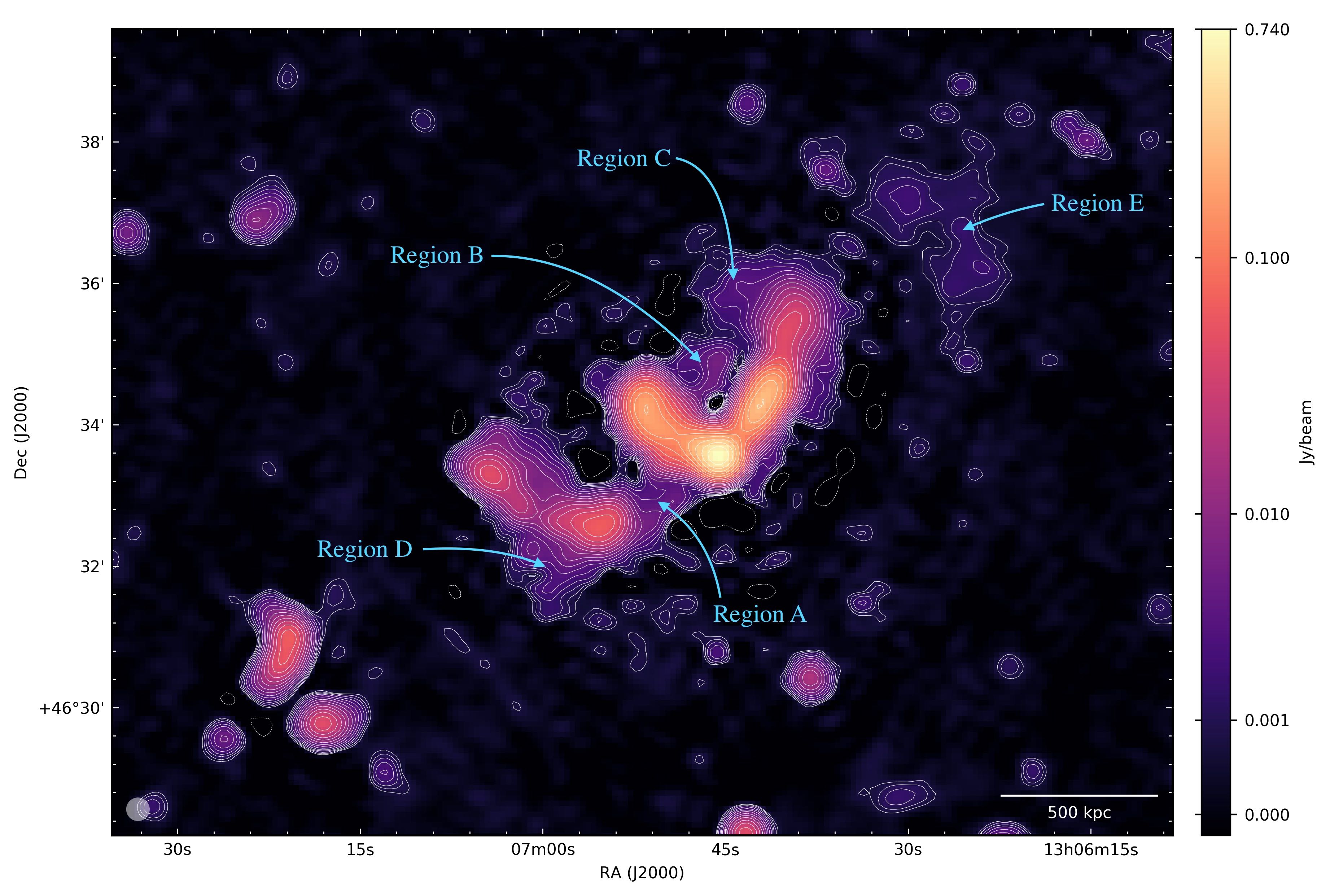}
\caption{Colour-scale image of A1682 on a log stretch from $-130~\mu \textrm{Jy/beam}$ to 0.74~Jy/beam (the brightest pixel is 0.74 Jy/beam). The RMS noise is $130~\mu \textrm{Jy/beam}$. Dashed contours are overlaid at -5 times the RMS noise. Solid contours start at 5 times the RMS noise and increase by factor 1.4. The resolution is $20\arcsec \times 20\arcsec$.}
\label{figure:lofarlowres}
\end{figure*}

\begin{figure*}
\includegraphics[width=\hsize]{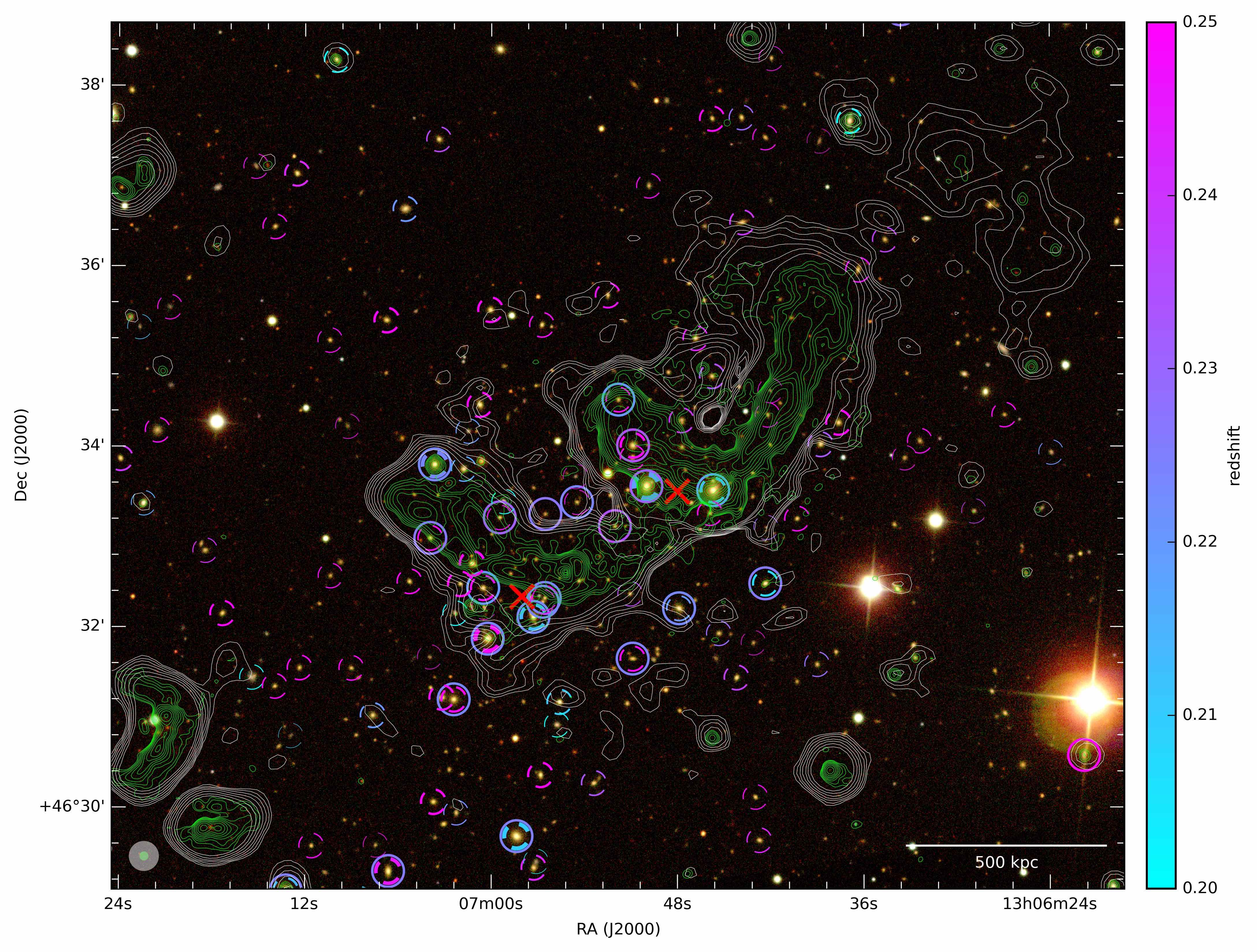}
\caption{SDSS background composite image from bands \textit{g}, \textit{r} and \textit{i}. LOFAR contours at 20\arcsec resolution are overlaid in white starting at 5 times the RMS noise and increasing by a factor 1.4. LOFAR contours at 6\arcsec\, resolution are overlaid in green starting at 5 times the RMS noise and increasing by a factor 1.6. Redshift coloured markers are the same as Figure \ref{figure:galdistribution}. Red crosses mark the centres of mass of the two sub-clusters identified by \protect \cite{dahle2002}.}
\label{figure:lofarall}
\end{figure*}

\begin{figure*}
\includegraphics[width=0.49\hsize]{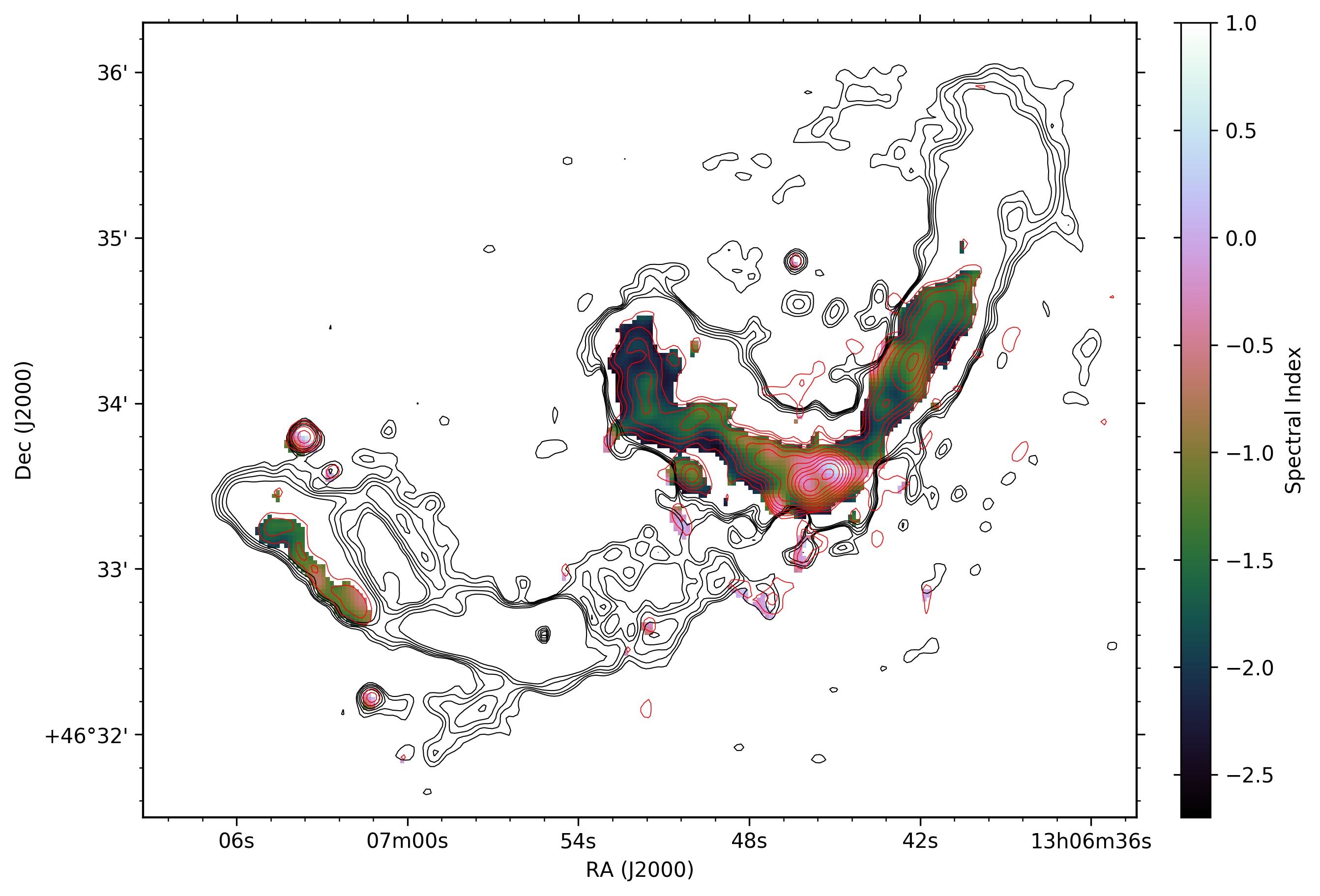}
\includegraphics[width=0.49\hsize]{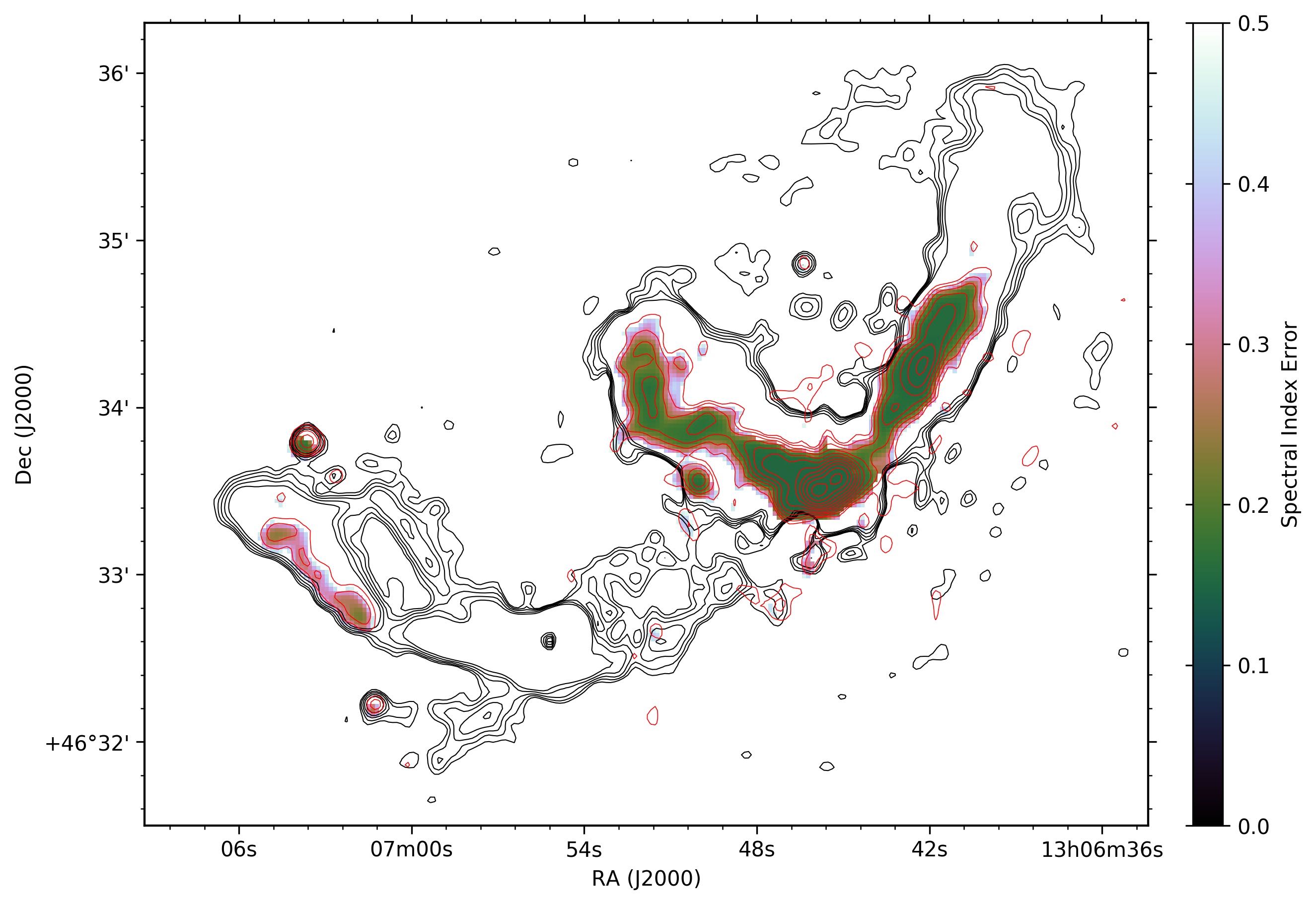}
\caption{Left: Spectral index map from our reprocessed GMRT data at 610~MHz to our LOFAR HBA data at 150~MHz. Both maps are at a resolution of $6\arcsec \times 6\arcsec$. Contours start at 5 times the RMS noise ($90~\mu \mathrm{Jy/beam}$ for both LOFAR and GMRT), and increase by a factor 1.7. Right: The corresponding spectral index error map for each pixel.}
\label{figure:spectralindexmap}
\end{figure*}

\begin{figure*}
\includegraphics[width=1.0\hsize]{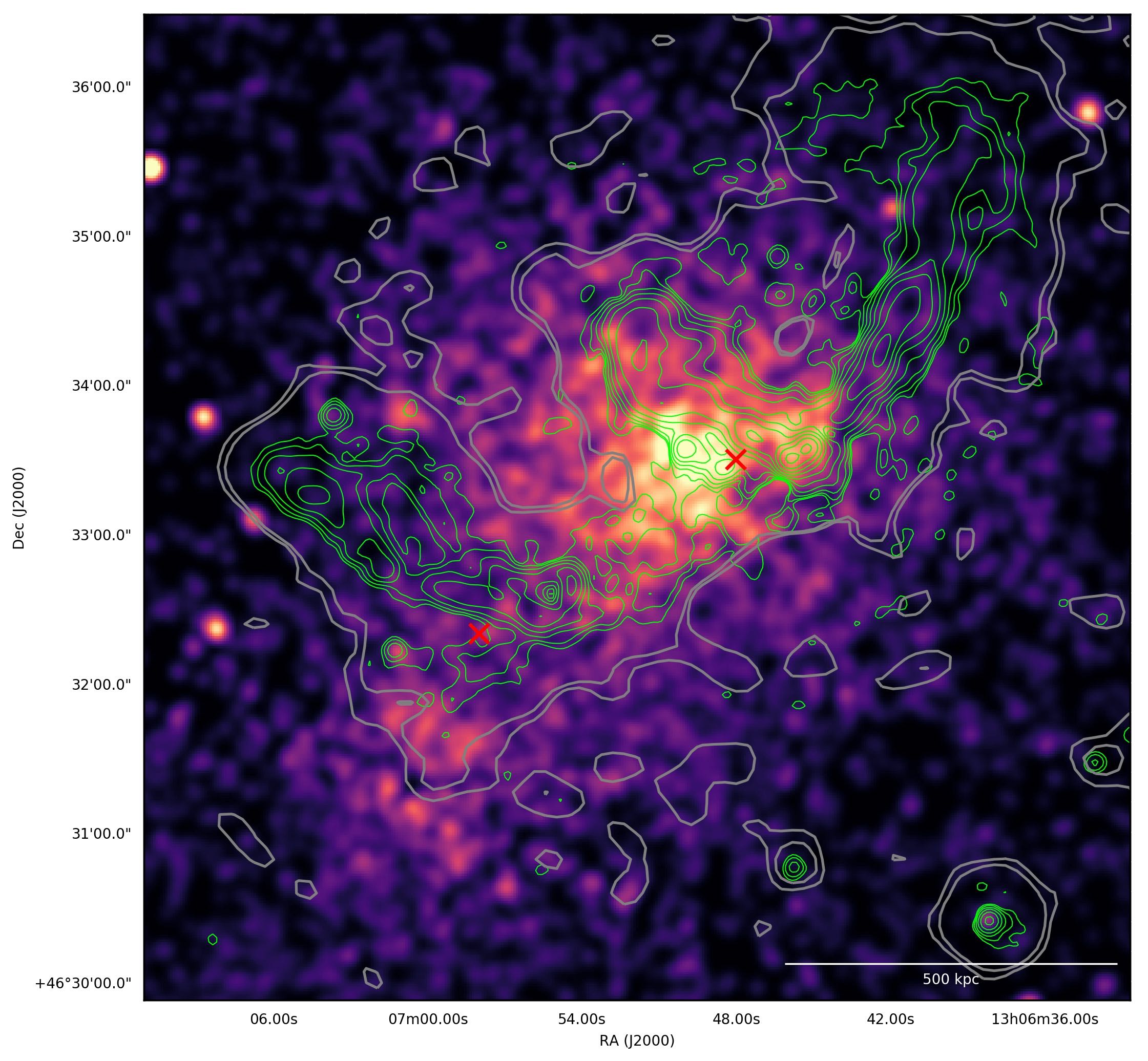}
\caption{X-ray mosaic from Chandra (ObsID's: 3244 and 11725) in the 0.5-2.0 keV band, smoothed with a Gaussian kernel of standard deviation $5\arcsec$. LOFAR contours are overlaid in green starting at 5 times the RMS noise ($90~\mu \textrm{Jy/beam}$) and increase by factor 2, at a resolution of $6\arcsec \times 6\arcsec$. Grey contours shows LOFAR data at $20\arcsec$ resolution at 5 and 10 times the RMS noise ($130~\mu \textrm{Jy/beam}$). Red crosses mark the centre of mass of each of the sub-clusters as in other figures.}
\label{figure:xray}
\end{figure*}

\begin{figure*}
\includegraphics[width=0.48\hsize]{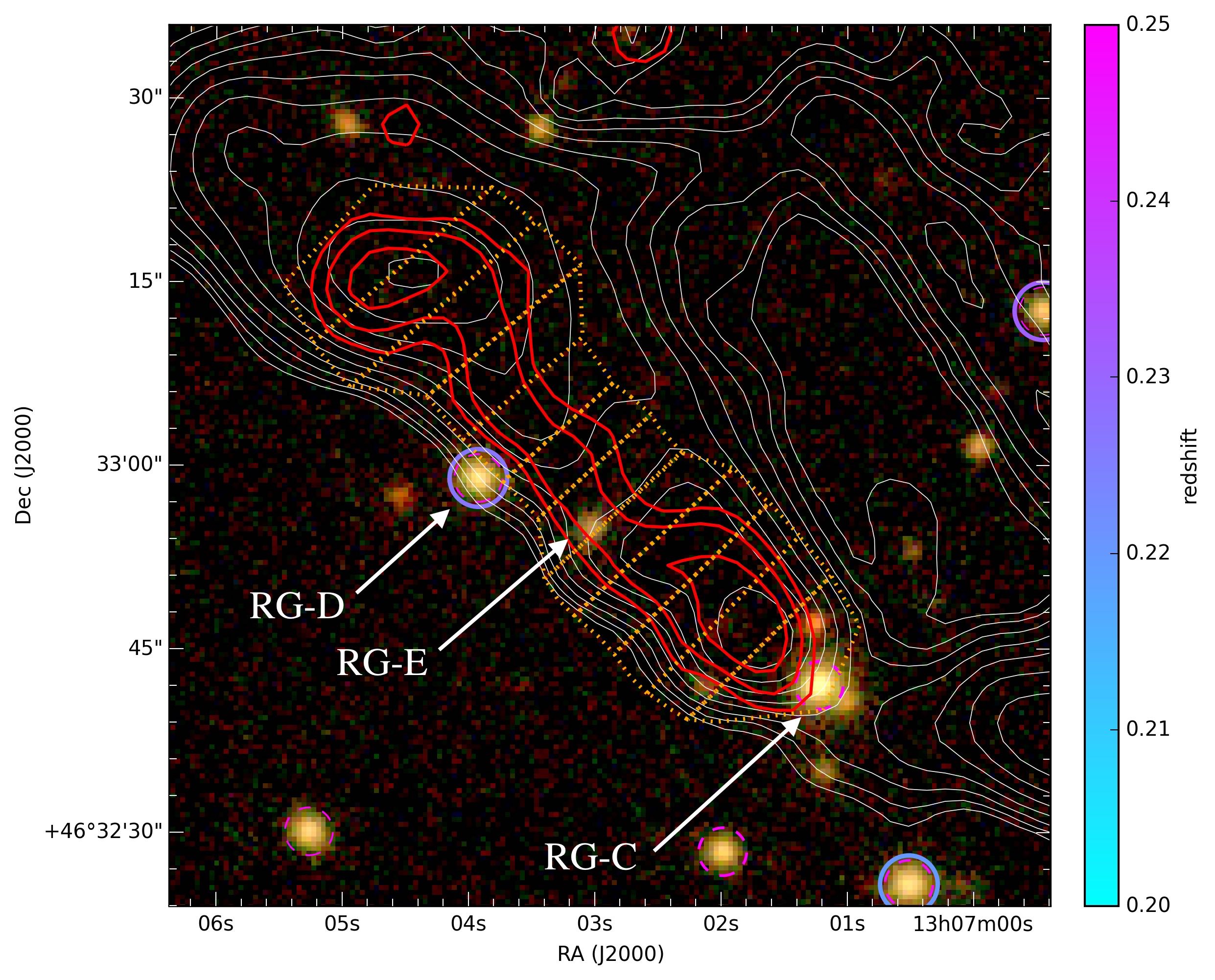}
\includegraphics[width=0.50\hsize]{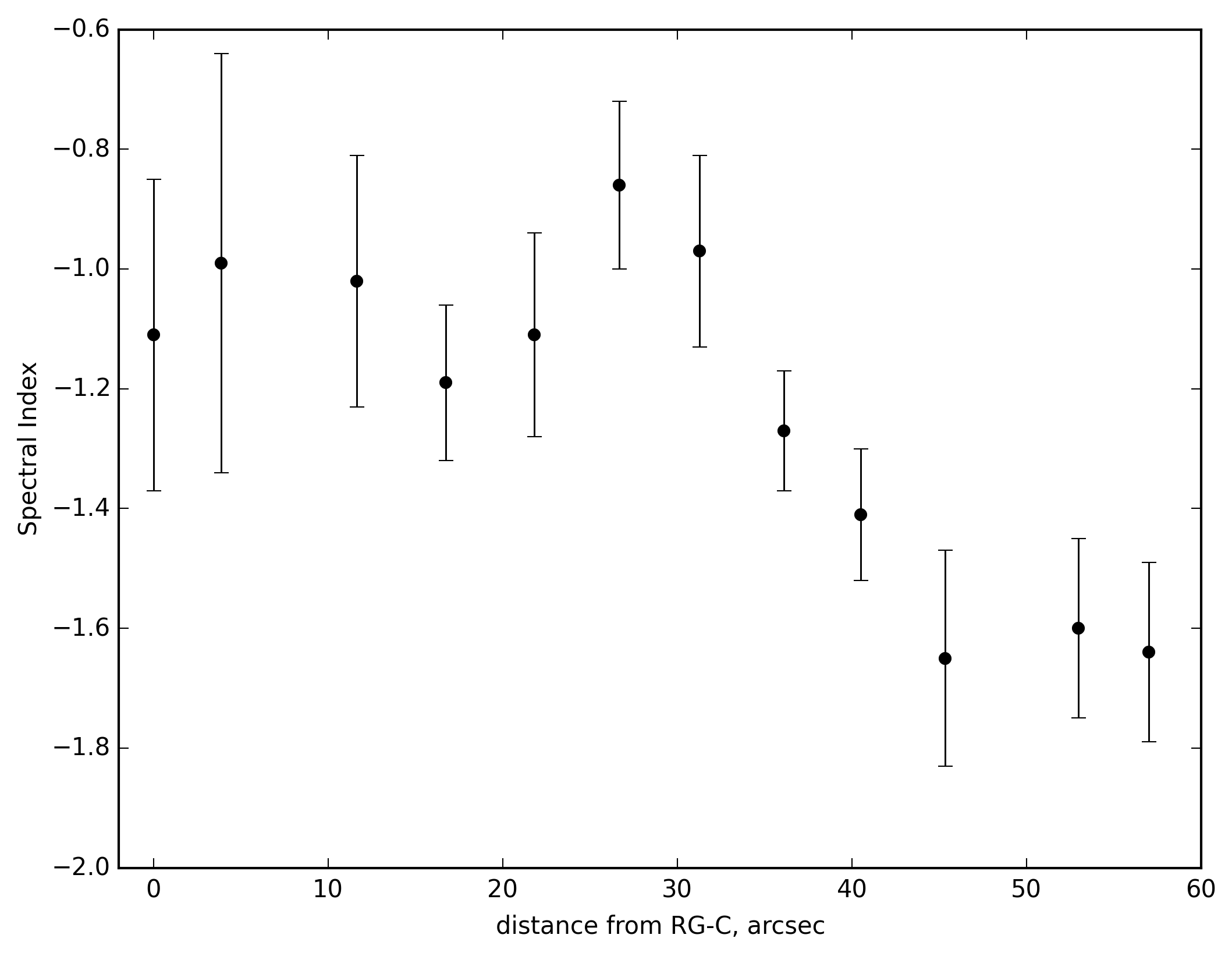}
\caption{Left: Zoom in on radio galaxy candidates RG-C, -D and -E. LOFAR contours are in white, GMRT contours in red, both at 6\arcsec\, resolution starting at 5 times the RMS noise ($90~\mu \mathrm{Jy/beam}$) and increasing by factor 1.4. Orange dashed lines mark the position of spectral index measurements. Right: Spectral index as a function of distance from RG-C. Measurements are an average of all pixels in the region, taking all emission above 5 times the RMS noise.}
\label{figure:RG-D}
\end{figure*}


\section{Analysis}\label{section:analysis}
Throughout the cluster we observe complex radio emission with LOFAR at high and low resolution. The total integrated flux density (all emission above 5 times the RMS noise) of Abell 1682 at 150~MHz is $3.53 \pm 0.71 \,\mathrm{Jy}$ in both the 6\arcsec\, and 20\arcsec\, images (Figures \ref{figure:lofar}, \ref{figure:lofarlowres}). It is challenging to disentangle contributions from the ICM, the BCGs, or other AGN throughout the cluster. Here we discuss potential origins of each component considering ancillary radio, X-ray and optical data, referring closely to Figures \ref{figure:lofar}, \ref{figure:lofarlowres}, \ref{figure:lofarall}, \ref{figure:spectralindexmap}, \ref{figure:xray}, \ref{figure:RG-D} and sources labelled within.

\subsection{AGN, radio jets and lobes}
Figure \ref{figure:lofar} presents our 6\arcsec\, resolution LOFAR data showing the detailed morphology of the radio structures in this galaxy cluster. In these new data, BCG-A shows an extension of approximately 30\arcsec\, to the south-west (towards the red cross marking the sub-clusters centre of mass) resembling a radio jet from its AGN that has not been seen in previous data. However, we cannot rule out that this extension is a result of a different source of radio emission in this complex area of the cluster (another AGN), or has contamination from a background high redshift AGN. BCG-A is coincident with the peak in the X-ray image (Figure \ref{figure:xray}), and whilst it hosts a radio loud AGN capable of producing large radio jets, it is currently not clearly connected to the surrounding extended radio structures. BCG-A has a peak flux density of $27 \pm 4 \mathrm{mJy/beam}$ at 150~MHz, and fitting a single component Gaussian (ignoring the extension to the south-west) gives an integrated flux density of $44 \pm 7 \mathrm{mJy}$ at 150~MHz.

A prominent optical source in this region is labeled as RG-A, which could be a tailed radio source associated with the emission highlighted by the dotted green line (resembling a wide-angle tailed radio galaxy moving south-east). The spectral index in the region around RG-A is $\alpha = -1.70 \pm 0.19$ (150--610~MHz, see Figure \ref{figure:spectralindexmap}). BCG-A has an average spectral index of $\alpha = -1.51 \pm 0.29$ calculated over the island of 74 pixels in Figure \ref{figure:spectralindexmap}. The lack of any flatter spectrum components aligned with RG-A or BCG-A suggest neither currently have active radio jets. If either BCG-A or RG-A are responsible for the extended radio emission in this area, regions B and C in Figure \ref{figure:lofarlowres} could be emission from earlier in their lifetime when they were position farther to the north-west. Alternatively, these regions could be a result of the radio lobes expanding out into less dense regions of the intra-cluster medium.

We note that to the south-east of RG-A the radio emission extends to overlap a large optical source, however this source is classified as a star in SDSS data. Our spectral index map indicates the emission flattens in this area, but the error is very large and we conclude it is unlikely there are any other optical counter-parts responsible for this emission other than BCG-A, RG-A, or BCG-B.

BCG-B has a peak flux density of 0.29 Jy/beam in Figure \ref{figure:lofar}, and the component 1.2\arcsec\, to the north-west of this (likely to be where its jet has left the galaxy) has a peak flux density of 0.28 Jy/beam. The average spectral index on top of BCG-B is $\alpha = -0.65 \pm 0.15$ (averaged over 56 pixels centred on BCG-B which are above 1000 times the RMS noise in the LOFAR image). There is a steady fall off in surface brightness going from BCG-B to the south-west, with little change in the perpendicular direction (north-west to south-east). This agrees with a scenario in which BCG-B is moving north-east with its radio jets oriented perpendicular to this direction of travel (as shown in \cite{venturi2013}). As such, at an earlier point in time the alignment of BCG-B could explain the resolved structures to the south-west of BCG-B in Figure \ref{figure:lofar}. However some of this emission is seen spuriously in the GMRT image at 610~MHz (Figure \ref{figure:spectralindexmap}) and is spatially coincident with a galaxy directly to the south of BCG-B with photometric redshift (indicating it is a cluster member). The detection is not significant enough to obtain a reliable spectral index estimation. Optical sources surrounding BCG-B are also likely to be cluster members and could be contributing to the radio emission with AGN of their own.

In the south-east of the cluster the emission consists of numerous AGN identified as cluster members and two regions of steep spectrum emission previously labelled as the S-E ridge and the diffuse component. The S-E ridge is resolved in these new data, and is likely associated with RG-C, -D, or -E in Figure \ref{figure:lofar}. Whilst RG-D is the only one with spectra in SDSS, RG-C and -E are very likely to be cluster members and have photometric redshifts of $0.254 \pm 0.021$ and $0.312 \pm 0.110$ respectively. The spectral index information presented in Figures \ref{figure:spectralindexmap} and \ref{figure:RG-D} indicate that the host is in the centre of this structure where the spectral index is flattest ($\alpha = -0.83 \pm 0.12$). The spectrum of the emission then steepens either side of this (which is expected as the electrons age), reaching $\alpha = -1.64 \pm 0.15$ on the north-east side, and $\alpha = -1.11 \pm 0.26$ on the south-west side. The asymmetry of the spectral index profile around the central optical counterpart could be caused by differing density of the ICM, as seen in studies such as \cite{gasperin2017}. Given the spectral index profile and morphology of the radio emission, the most likely optical counterpart is RG-D or -E in the centre where the spectrum is flattest, although RG-D is slightly to the south-east of the GMRT emission. 

While RG-C appears as possible candidate in terms of its spatial location, the spectral index profile of the extended radio emission and lack of a flatter spectrum component near RG-C suggests it is unlikely to be the cause. There is a drop in the brightness of the LOFAR emission around RG-C as it connects to the diffuse component in the south-west. Furthermore, no GMRT emission (indicative of a counter-jet) is seen to the south-west of RG-C which would have been detected if it were of similar spectral index values. Therefore, emission to the south-west of RG-C appears to become significantly steeper. Whilst RG-C could potentially be the host of a large radio galaxy contributing to both the S-E ridge and diffuse component, overall we conclude that the S-E ridge is likely of a separate origin to the diffuse component that we discuss in Section \ref{section:diffuse_component}.

\subsection{The N-W ridge: a radio relic or lobe?}
The extended structure of the N-W ridge resembles the jet and lobe of a radio galaxy originating from a powerful AGN in BCG-B (green lines in Figure \ref{figure:lofar}). The orientation of BCG-B's jets from the higher resolution VLA observation \citep{venturi2013} agree with this scenario. Furthermore, the AGN at the centre of BCG-B could be responsible for all the emission traced by green lines in Figure \ref{figure:lofar}. Our spectral index map from this LOFAR 150~MHz observation to our reprocessed GMRT observation at 610~MHz (Figure \ref{figure:spectralindexmap}) shows flat spectrum emission around BCG-B ($\alpha = -0.65 \pm 0.15$), which becomes steep spectrum emission in both directions traced by green lines in Figure \ref{figure:lofar} ($\alpha = -2.1 \pm 0.2$ in the east lobe and $\alpha = -1.96 \pm 0.16$ in the N-W ridge), which is typical of a large radio galaxy with lobes extending hundreds of kpc. However, there is also the possibility that the N-W ridge is a radio relic, or has at least been partially re-accelerated by the cluster merger.

The spectral index information from our LOFAR 150~MHz observation to our reprocessed GMRT observation at 610~MHz (Figure \ref{figure:spectralindexmap}) shows very little variation along the length and width of the N-W ridge. The steepest part is at the start of the ridge near BCG-B with a spectral index of $\alpha = -1.96 \pm 0.16$. In the middle of the ridge the spectral index flattens to $\alpha = -1.1 \pm 0.15$ (which is coincident with various optical counter-parts in Figure \ref{figure:lofarall} that could host AGN). Near the end of the ridge farthest from BCG-B, the spectral index is $\alpha = -1.45 \pm 0.17$. Overall, we see that the N-W ridge is very steep spectrum emission, without any significant gradient along its length other than when it connects with BCG-B, becoming $\alpha = -0.65 \pm 0.15$. This jump to flatter spectral index values near BCG-B indicates its AGN has more recently restarted. Whilst we do not observe a spectral index gradient perpendicular to the length of the ridge either (as seen in \citealt{venturi2013}), we do see the LOFAR emission at 150~MHz extends farther south-west than the GMRT emission at 610~MHz, indicating the spectrum steepens in this direction. If the N-W ridge is a radio relic this extra steep spectrum emission seen by LOFAR would be expected if formed from a shock moving north-east, as proposed in \cite{venturi2013}. However, the same effect could also be achieved by the radio lobe expanding to the east, or if BCG-B has moved north-east with its radio lobe, leaving older and steeper spectrum emission to the south-west of the N-W ridge as we see with LOFAR.

If the N-W ridge was re-accelerated in a radio relic like scenario, its position is a-typical, being perpendicular to the cluster merger axis. In a simplified geometrical set up assuming the relic is viewed side-on, it would be expected that a shock front is moving from south-west to north-east, which is counter-intuitive to the cluster merger axis of south-east to north-west. Furthermore, compressed, denser X-ray gas should be found to the south-west. In the X-ray image (Figure \ref{figure:xray}) there is no indication of a surface brightness jump coinciding with the N-W ridge. In simulations of radio relic formation there are cases of a-typical relics in complex merger scenarios, such as object A in Figure 7 of \cite{nuza2017}. The N-W ridge in A1682 is similar to this, and it could be the case that A1682 is one of the few examples of relics which are not caused by a peripheral, spherical shock. 

The N-W ridge also extends hundreds of kpc farther to the north-west than previous observations, and with the non-detection in the GMRT data in this region, we infer this is an area of very steep spectrum emission, constraining the spectral index as $\alpha < -2.5$ given the non-detection with the GMRT. Given this emission reaches up to 0.6 Mpc from BCG-B, if it originated from BCG-B it is likely to be very old emission with a very steep spectral index. 

Overall, we believe the most likely origin of the N-W ridge is emission from a previous epoch of AGN activity (an old radio lobe) associated with BCG-B. The jump in spectral index from steeper to flatter values around BCG-B it is likely caused by the AGN in BCG-B having more recently restarted. This emission over the N-W ridge has the potential to have been re-accelerated by the merger event in the region closer to the centre of the cluster, although we find no conclusive evidence given this data.

\subsection{The diffuse component} \label{section:diffuse_component}
The diffuse component identified in previous studies as an USSRH candidate is resolved in this new observation as a feature distinct from the S-E ridge. The resolved radio emission traced by the solid cyan lines in Figure \ref{figure:lofar} is morphologically indicative of a wide-angle tailed \citep[WAT;][]{sakelliou2000} radio galaxy. A likely optical counter-part identified in SDSS is RG-B (13:06:54.34, +46:32:30.46). If so, this source could be moving south-west where its radio-loud AGN has left the observed radio structures in its wake. The extension of the radio lobes to the north-east could also be due to the jet propagation at relativistic speeds and the cluster environment influencing their morphology. RG-B looks to be a merging galaxy, and has a photometric redshift in SDSS of $0.293 \pm 0.063$. Despite the general bias towards higher redshifts for photometrically observed galaxies it is highly likely to be a cluster member. The region where the two cyan lines join to create the very low brightness tail (traced by a dashed cyan line) could represent a period where the radio jets from the AGN were of much lower power or have been more confined by the ICM, only creating a single traced line of emission \citep[often referred to as narrow-angle tailed radio galaxies;][]{stocke1987}.

If the emission highlighted by cyan lines is indeed associated with RG-B then its spectral properties in the radio emission make it difficult to deduce a precise evolutionary scenario. Previous studies show this emission has a very steep spectrum, with none of the emission detected by \cite{venturi2008} in their deep GMRT observations at 610~MHz. Given our LOFAR observations and their non detection we expect a spectral index of $\alpha < -2.5$. If RG-B is currently (or has recently been) active, it is expected that emission near the proposed optical host should have a much flatter spectrum. It therefore appears clear that the AGN in RG-B is no longer radio-loud and has been inactive for a significant period of time. Such sources are referred to as remnant radio galaxies which typically have ages around a few hundred million years \citep[e.g.][]{brienza2017, Murgia2011}. It therefore seems most likely that RG-B has been sitting more or less where it currently is for approximately this period of time since it was last active, and has not moved from north-west to south-east, unless its motion has been halted in its current location for a significant period of time. Instead, the propagation of the jets, along with the influence from the intra-cluster medium, has created the morphology of the radio structures observed in this area.

Whilst RG-B is the most compelling candidate for the diffuse component in terms of its location, there are numerous optical counter-parts in the region (all are likely cluster members) which could also have contributed to the radio emission. If the host was moving from north-east to south-west, it is likely to be one of the sources farther to the south-west where there is no observed radio emission, and not RG-B. 
 
Overall, the very steep spectrum emission in this area of the cluster is likely a result of significant synchrotron ageing from very old radio galaxy emission. Within a galaxy cluster it has been proposed that remnant radio sources can be observed for longer than would be expected due to lower adiabatic losses since the dense cluster environment prevents the radio lobes expanding as much \citep{Murgia2011}. It could also be the case that re-acceleration mechanisms as a result of the cluster merger have also contributed to highlighting this remnant source with such a steep spectrum. However, without detailed modelling over a wide bandwidth \citep[such as analysis done by][]{gasperinGREET2017}, it is not possible to distinguish between ageing electrons in an old radio galaxy tail, and re-acceleration.

\subsection{Searching for an USSRH}
In previous studies the emission labeled as the diffuse component was classified as a potential USSRH, but as discussed our high resolution LOFAR images show that this is unlikely. Instead we investigate the area between the diffuse component and the BCGs as a potential USSRH (region A in Figure \ref{figure:lofarlowres}). This area is closer to the centre of the X-ray emission from the cluster (see Figure \ref{figure:xray}), where re-acceleration mechanisms as a result of the merger event will be more powerful. The low resolution of previous spectral studies do not distinguish this area from the diffuse component. Our LOFAR observations show the presence of significant residuals in this area at 6\arcsec\, resolution (see Figures \ref{figure:lofar} and \ref{figure:lofarall}), and direct detection of significantly extended emission at 20\arcsec\, resolution (see Figures \ref{figure:lofarlowres} and \ref{figure:lofarall}). This suggests that particle acceleration has occurred in this area over large spatial scales (up to 500 kpc). The diffuse and low surface brightness emission encompassing region A is unlikely to be connected with any AGN identified in the high resolution image, and is in a location between the two sub-clusters (close to the centre of the X-ray emission; Figure \ref{figure:xray}) which supports an origin from the re-acceleration of electrons in the wake of a merger event.

We re-imaged the 610~MHz GMRT data at a resolution of 20\arcsec\, (the same as our LOFAR map in Figure \ref{figure:lofarlowres}) reaching an RMS noise of $280\mu\mathrm{Jy}$, showing no emission in region A. Given the non-detection of emission in this region at 610~MHz, we show this is relatively steep spectrum emission with a spectral index of $\alpha < -1.1$, which is a signature of re-acceleration models \citep{brunetti2008nat,brunetti2014}. However, without spectral modelling over a wide bandwidth we cannot tell between re-acceleration and the ageing of fossil radio emitting plasma which could be left over from old radio galaxies \citep{ensslin2001}. Overall we conclude that region A is very likely to be a radio halo caused by the merger event. The presence of many AGN with radio jets throughout the cluster present an obvious mechanism to fill the ICM with a seed population of relativistic electrons to fuel this radio halo emission.

\subsection{Additional diffuse emission}
Other areas of extended diffuse emission are located around the cluster, as labelled in Figure \ref{figure:lofarlowres}. Whilst region A is the strongest, regions B, C and D also show significant diffuse emission not seen before and without significant detection in the 6\arcsec\, image. These regions could be very old and mildly relativistic electrons left over from the lobes of radio galaxies, with the possibility that they have been re-accelerated by the cluster merger event. Region B could be old emission from the lobes of a radio galaxy moving south-east, associated with RG-A or BCG-A and the dashed green lines in Figure \ref{figure:lofar}. Region C could also be associated with this, but appears to be disconnected, so instead could be an extension of the N-W ridge. In region D there are numerous radio loud galaxies and candidates (shown in Figure \ref{figure:lofarall}) which could have had radio lobes that contributed to this emission.

Region E is distinct from the rest of the cluster. It shows minimal emission in the 6\arcsec\, resolution image, and is only significantly detected in the 20\arcsec\, resolution image. There is one photometrically identified cluster member to the north-east of this region with a radio loud AGN (dashed cyan circle in Figure \ref{figure:lofarall}), though it is disconnected from region E. There is a bright galaxy in the centre of region E with the appearance of an orange elliptical (13:06:27.84, +46:36:40.14) which has a photometric redshift of $0.302 \pm 0.047$. It has a slightly disturbed morphology indicating it is undergoing a merger event. This could be the host of a remnant radio galaxy currently with no radio-loud AGN, where its lobes have aged becoming very faint and diffuse. Its extended radio structure is measured at $154 \pm 10$ arc-seconds, giving this giant remnant radio galaxy a physical size of $1.29 \pm 0.29 \,\mathrm{Mpc}$ (where the error in the photometric redshift dominates). It has an integrated flux density of $30.6 \pm 6.1 \,\mathrm{mJy}$, giving a luminosity of $9.61 \substack{+6.49 \\ -2.40} \times 10^{24} \,\mathrm{W/Hz}$ at 150~MHz. Given the observed bias towards higher values for photometric redshifts of cluster members, we also calculate its physical size based on the average redshift of the cluster (z=0.225), as $0.86 \pm 0.06 \,\mathrm{Mpc}$, giving a luminosity of $2.83 \pm 1.02 \times 10^{24} \,\mathrm{W/Hz}$. Whilst region E is elongated and aligned perpendicular to the cluster merger axis and morphologically indicative of a radio relic, it is more than a Mpc away from the cluster centre at the clusters redshift (and with an optical host at a higher redshift), making this it very unlikely to be a relic.

\subsection{Giant radio galaxies}
There are numerous extended radio sources observed surrounding the galaxy cluster which are likely background sources (AGN with radio jets and lobes) at higher redshift. Many of them do not have discernible optical counter-parts in the SDSS image. However, we note a handful of them which do and present them as giant radio galaxies with both Fanaroff-Riley type I and II \citep[FR-I, FR-II;][]{FR1974} morphology.

\begin{figure*}
\includegraphics[width=0.5\hsize]{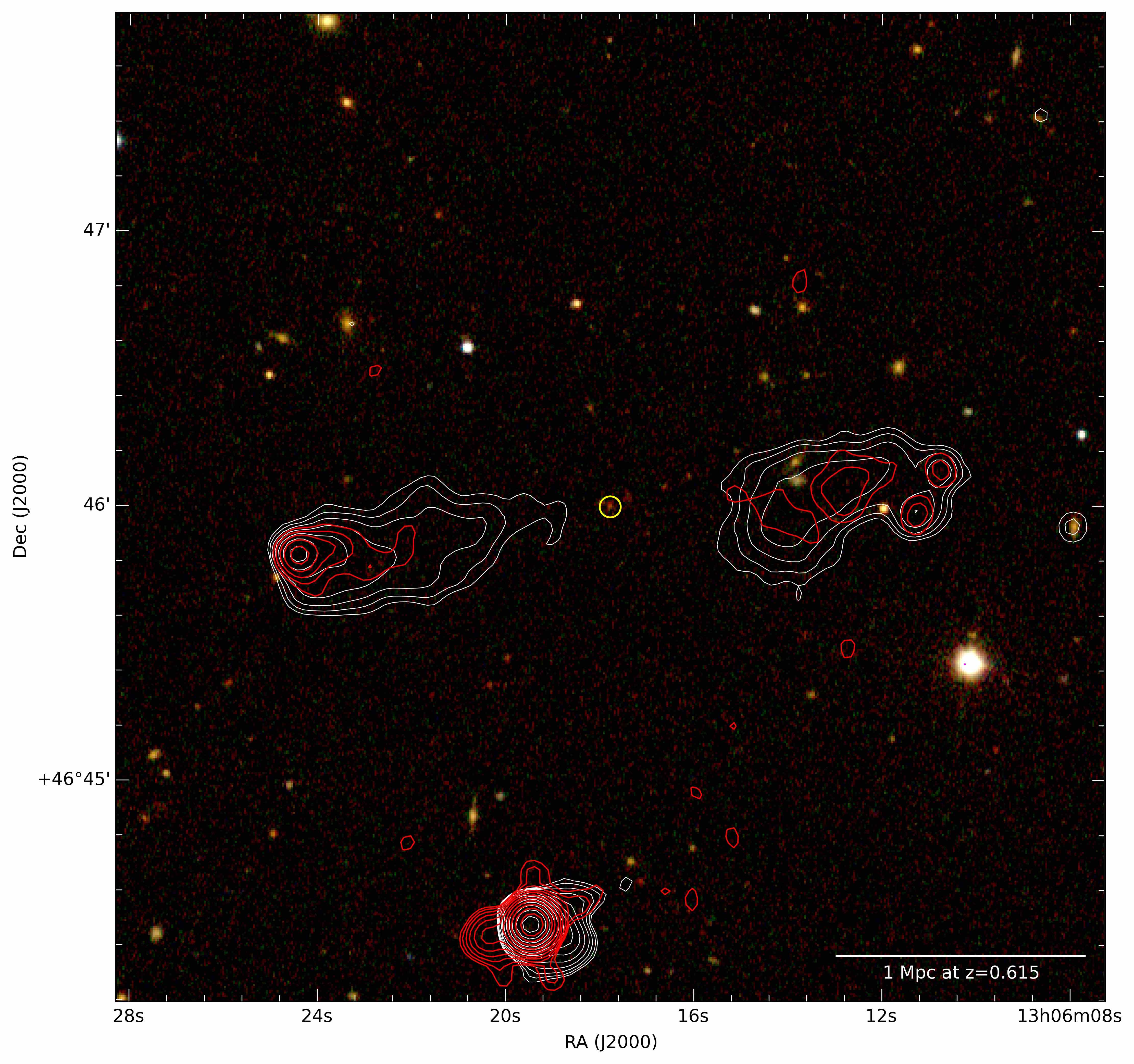}
\includegraphics[width=0.5\hsize]{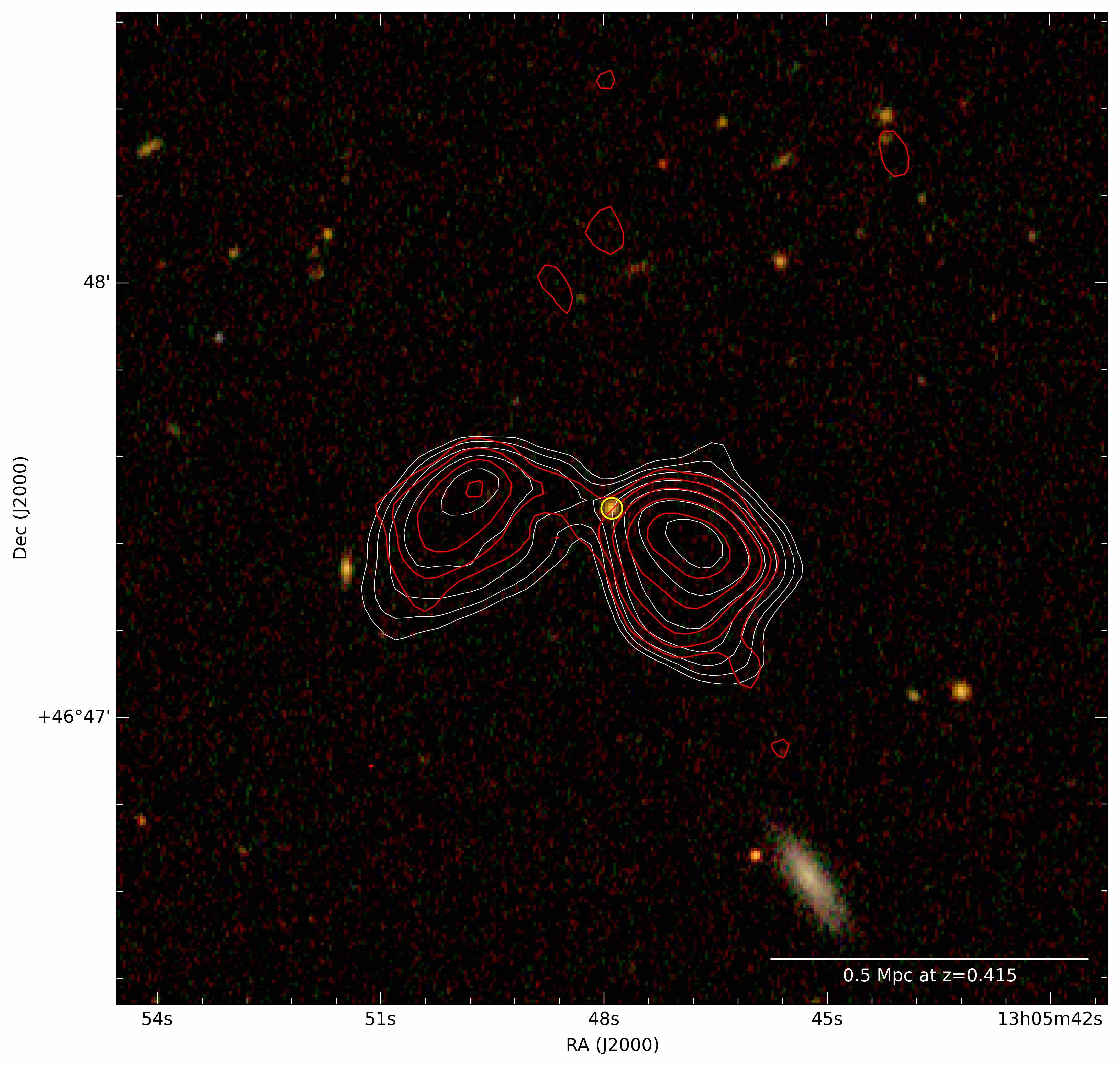}
\caption{Left: A giant radio galaxy with the likely optical host circled in yellow, identified in SDSS (13:06:17.78, +46:45:59.77) with a photometric redshift of $0.615 \pm 0.060$. This giant radio galaxy has a physical size of $2.86 \pm 0.55  \,\mathrm{Mpc}$. Right: A giant radio galaxy with the likely optical host circled in yellow, identified in SDSS (13:05:47.89, +46:47:28.85) with a photometric redshift of $0.415 \pm 0.035$. This radio galaxy a physical size of $0.68 \pm 0.20 \,\mathrm{Mpc}$. In both images the background is an SDSS composite image from bands \textit{g}, \textit{r} and \textit{i}. LOFAR and GMRT contours are overlaid at $6\arcsec\,$ resolution starting at 5 times the RMS noise ($90~\mu \mathrm{Jy/beam}$), and increase by a factor 1.6.}
\label{figure:RG-1}
\end{figure*}

In Figure \ref{figure:RG-1} we present two new radio galaxies to the north of the cluster. The radio galaxy in the left of this figure has an optical host identified in SDSS (13:06:17.78, +46:45:59.77) with an SDSS \textit{r}-band magnitude of 21.86 and a photometric redshift of $0.615 \pm 0.060$. The extended radio structure is measured at $156 \pm 10$ arc-seconds, giving this giant radio galaxy a physical size of $2.86 \pm 0.55  \,\mathrm{Mpc}$ (combined errors from angular size and redshift). There are three other other galaxies situated between, and in line with, the radio lobes which are less likely candidates as optical hosts but could be part of a galaxy group (\textit{a}: 13:06:17.45, +46:46:01.59, \textit{b}: 13:06:16.64, +46:46:04.31, \textit{c}: 13:06:16.12, +46:46:06.66). They lie 12, 18 and 25 arc-seconds the west of the proposed optical host, respectively, with SDSS \textit{r}-band magnitudes of $24.41 \pm 0.54$, $22.98 \pm 0.20$, and $22.96 \pm 0.30$. Candidates \textit{a} and \textit{c} have photometric redshifts of $0.879 \pm 0.1443$ and $0.696 \pm 0.1108$ respectively. This giant radio galaxy has a distinct FR-II morphology with two separate lobes. The left lobe has an integrated flux density of $47.5 \pm 9.5 \,\mathrm{mJy}$ at 150~MHz, and $8.5 \pm 1.7 \,\mathrm{mJy}$ at 610~MHz. The average spectral index of the left lobe is $\alpha = -0.74 \pm 0.32$ with a standard deviation of 0.32. The right lobe has an integrated flux density of $43.9 \pm 8.8 \,\mathrm{mJy}$ at 150~MHz, and $8.0 \pm 1.6 \,\mathrm{mJy}$ at 610~MHz. The average spectral index of the right lobe is $\alpha = -0.74 \pm 0.37$ with a standard deviation of 0.24. In total its luminosity is $1.56 \substack{+0.79 \\ -0.58} \times 10^{26} \,\mathrm{W/Hz}$ at 150~MHz.

The radio galaxy in the right of Figure \ref{figure:RG-1} has an optical host identified in SDSS (13:05:47.89, +46:47:28.85) with an SDSS \textit{r}-band magnitude of $20.34 \pm 0.03$, and photometric redshift of $0.415 \pm 0.035$. The extended radio structure is measured at $60 \pm 10$ arc-seconds, giving this radio galaxy a physical size of $0.68 \pm 0.20 \,\mathrm{Mpc}$. This radio galaxy has an FR-II morphology with two connected lobes. The left lobe has an integrated flux density of $23.9 \pm 4.8 \,\mathrm{mJy}$ at 150~MHz, and $10.0 \pm 2.0 \,\mathrm{mJy}$ at 610~MHz. The average spectral index of the left lobe is $\alpha = -0.50 \pm 0.31$ with a standard deviation of 0.24. The right lobe has an integrated flux density of $49.2 \pm 9.8 \,\mathrm{mJy}$ at 150~MHz, and $20.9 \pm 4.2 \,\mathrm{mJy}$ at 610~MHz. The average spectral index of the right lobe is $\alpha = -0.52 \pm 0.25$ with a standard deviation of 0.31. Considering the large size of the radio lobes these spectral index values are flatter than what would normally be expected, although they are consistent within the errors. In total its luminosity is $4.84 \substack{+2.20 \\ -1.70} \times 10^{25} \,\mathrm{W/Hz}$ at 150~MHz (where the error is dominated by the uncertainty in redshift).

\begin{figure*}
\includegraphics[width=0.5\hsize]{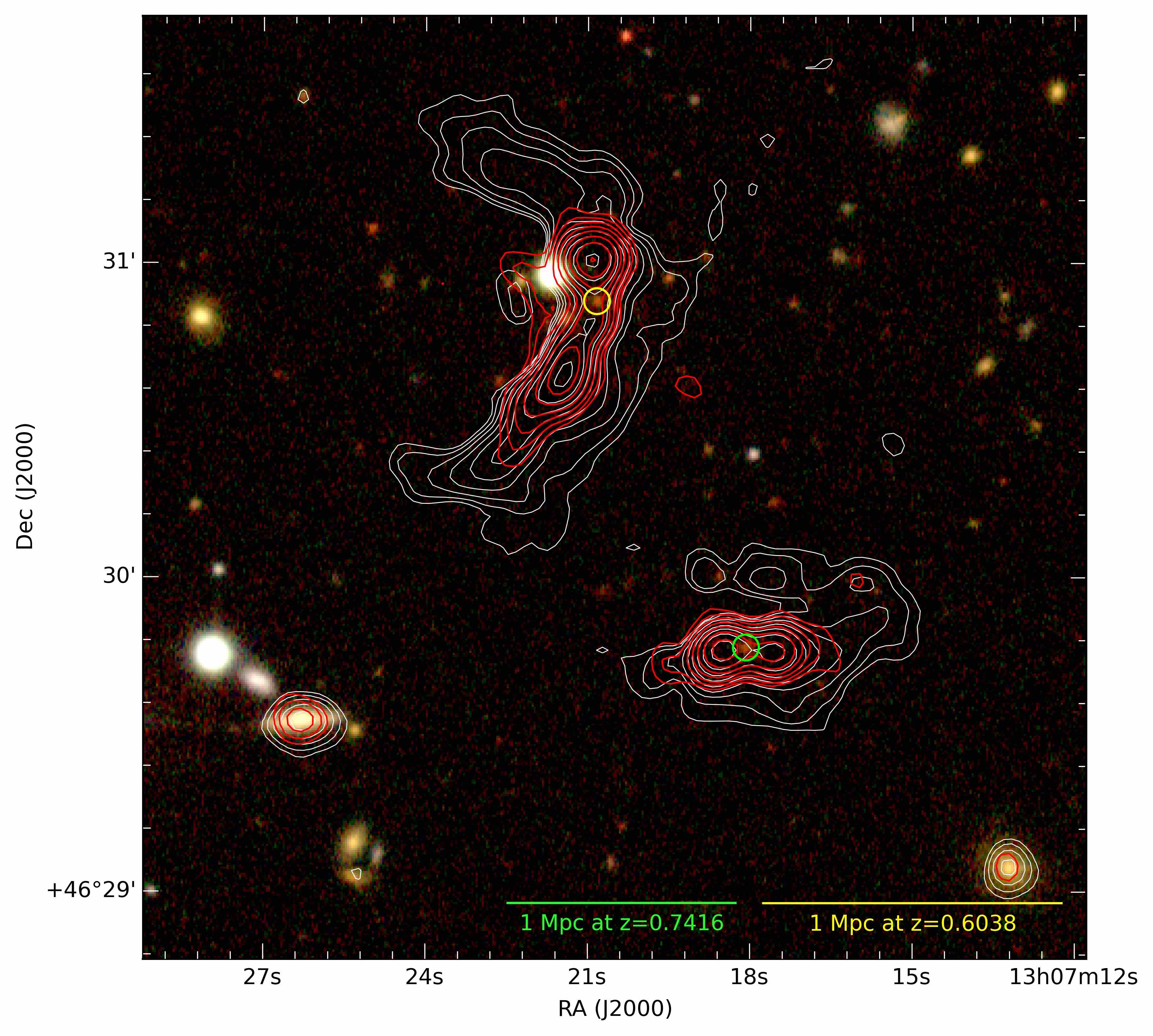}
\includegraphics[width=0.5\hsize]{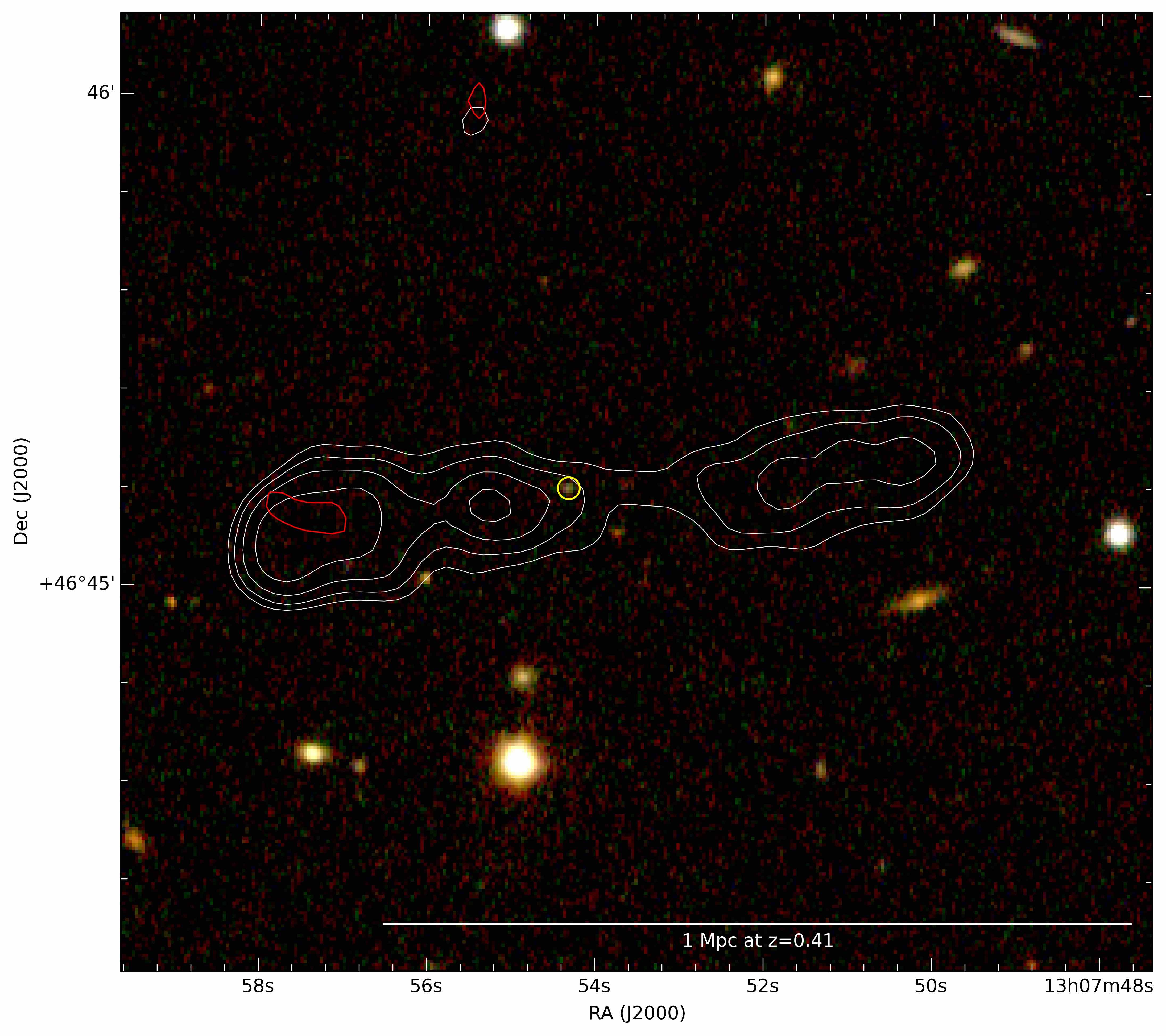}
\caption{Left: Two giant radio galaxies with likely optical hosts circled. In yellow (13:07:20.83, +46:30:52.67) the host galaxy has a spectroscopic redshift of $0.6038 \pm 0.0002$. This giant radio galaxy a physical size of $2.07 \pm 0.18 \,\mathrm{Mpc}$. In green (13:07:18.07, +46:29:46.61) the host galaxy has a spectroscopic redshift of $0.7416 \pm 0.0002$. This giant radio galaxy a physical size of $1.27 \pm 0.22 \,\mathrm{Mpc}$ measured from east to west. Right: A giant radio galaxy with a potential optical host circled in yellow, identified in SDSS (13:07:54.32, +46:45:11.92) with a photometric redshift of $0.409 \pm 0.104$. It has a physical size of $1.08 \pm 0.36 \,\mathrm{Mpc}$. In both images the background is an SDSS composite image from bands \textit{g}, \textit{r} and \textit{i}. LOFAR and GMRT contours are overlaid at $6\arcsec\,$ resolution starting at 5 times the RMS noise ($90~\mu \mathrm{Jy/beam}$), and increase by a factor 1.6.}
\label{figure:RG-2}
\end{figure*}

In the left of Figure \ref{figure:RG-2} there are two radio galaxies (also seen in the south-east of Figure \ref{figure:lofarall}) in the periphery of Abell 1682. The radio galaxy in the north of this figure is coincident with a bright foreground star, but has an optical host identified in SDSS (13:07:20.83, +46:30:52.67) that is well positioned in-line with the radio lobes (shown by a yellow circle). This host has an SDSS \textit{r}-band magnitude of $21.42 \pm 0.10$ and spectroscopic redshift of $0.6038 \pm 0.0002$. The extended radio structure is measured at $116 \pm 10$ arc-seconds, giving this giant radio galaxy a physical size of $2.07 \pm 0.18 \,\mathrm{Mpc}$ (where the error in angular size dominates and redshift error is negligible). There is another galaxy situated 5 arc-seconds to the south-east (13:07:21.43, +46:30:49.36) which is off-centre from the radio structure but is likely to be a part of the same galaxy group. This has a spectroscopic redshift of $0.6025 \pm 0.0002$ and an SDSS \textit{r}-band magnitude of $20.41\pm 0.08$. This giant radio galaxy has an FR-I morphology, with an integrated flux density of $186.1 \pm 28.0 \,\mathrm{mJy}$ at 150~MHz, and $46.3 \pm 9.3 \,\mathrm{mJy}$ at 610~MHz. This gives a luminosity of $3.04 \pm 0.61 \times 10^{26} \,\mathrm{W/Hz}$ at 150~MHz. The average spectral index is $\alpha = -0.80 \pm 0.23$ with a standard deviation of 0.47.

The radio galaxy in the south-west of Figure \ref{figure:RG-2} has an optical host identified in SDSS (13:07:18.07, +46:29:46.61) that is positioned in the centre of the radio lobes (shown by green circle). It has an SDSS \textit{r}-band magnitude of $20.82 \pm 0.08$ and spectroscopic redshift of $0.7416 \pm 0.0002$. The extended radio structure is measured at $55 \pm 10$ arc-seconds from east to west, giving this giant radio galaxy a physical size of $1.27 \pm 0.22 \,\mathrm{Mpc}$. This giant radio galaxy has an FR-I morphology, with an integrated flux density of $79.4 \pm 15.9 \,\mathrm{mJy}$ at 150~MHz, and $24.2 \pm 4.8 \,\mathrm{mJy}$ at 610~MHz. This gives a luminosity of $2.15 \pm 0.42 \times 10^{26} \,\mathrm{W/Hz}$ at 150~MHz. The average spectral index is $\alpha = -0.60 \pm 0.25$ with a standard deviation of 0.62.

In the right of Figure \ref{figure:RG-2} the only potential optical host with a redshift measurement in SDSS is circled in yellow (13:07:54.32, +46:45:11.92). Whilst it is slightly off-centre towards the east lobe, it is one of the most likely host galaxy candidate in the area. It has a photometric redshift of $0.409 \pm 0.104$, and an \textit{r}-band magnitude of $22.42 \pm 0.14$. There is another optical source 1 arc-second to the south-west just outside of the radio emission which has no redshift measurement. This source is classified as a star in SDSS, although it could well be a high redshift quasar. The extended radio structure is measured at $96 \pm 5$ arc-seconds, giving this giant radio galaxy a physical size of $1.08 \pm 0.36 \,\mathrm{Mpc}$ (where the error in redshift dominates). It has an FR-II morphology, where the left lobe has an integrated flux density of $24.3 \pm 4.9 \,\mathrm{mJy}$ at 150~MHz, and the right lobe has an integrated flux density of $12.1 \pm 2.4 \,\mathrm{mJy}$. In total this gives a luminosity of $2.33 \substack{+2.47 \\ -1.40} \times 10^{25} \,\mathrm{W/Hz}$ at 150~MHz (where the error is dominated by the uncertainty in redshift). There is a negligible detection with the GMRT at 610 MHz.

\section{Conclusions}\label{section:conclusions}
Deep radio observations at 150~MHz using LOFAR have built on previous knowledge to provide a clearer picture of the phenomena in Abell 1682 as well as detecting new emission not seen before. Given the complexity of the radio emission we detect, deducing a precise dynamical scenario for this cluster and the origin of all the emission is extremely difficult. 

At 6\arcsec\, resolution new filamentary structures are observed up to 600 kpc in length. Many of these are proposed to be the jets and lobes from old radio galaxies which are moving through the cluster. Both BCGs are seen to have extended radio features associated with them, with the small extension of BCG-A not been seen before. We present a radio spectral index map from $150-610$~MHz (through a re-analysis of archival GMRT data), showing that the emission is mainly steep spectrum (up to $\alpha = -2.1 \pm 0.2$), with flatter spectrum emission (with $\alpha = -0.65 \pm 0.20$) typical of AGN activity around both BCGs. The extended radio structure connected to BCG-B (labelled the N-W ridge) shows no gradient in the spectral index along its length or width, but is significantly more extended in our 150~MHz observations compared to observations at 610~MHz. Given the non-detection of this extension by archival observations at higher frequencies we place a limit on the spectral index of $\alpha < -2.5$. Its features and orientation are unlike a typical radio relic, however there is the possibility that the cluster merger event has contributed to the emission mechanisms by re-accelerating the electrons through a low Mach number shock. Overall it is most likely that the N-W ridge is an old radio lobe from AGN activity in BCG-B. Furthermore, the large radio structure extending from BCG-B to the north-east is likely to be the second lobe from the AGN in BCG-B.

The area in the south-east of the cluster previously identified as an USSRH is resolved into a component strongly resembling the tail from a radio galaxy moving south-east through the cluster, though this is mainly a morphological classification since there is no redshift for the optical counter part, or spectral index information on the host. Given that none of this emission is observed in archival GMRT observations at 610~MHz we place a limit on the spectral index of $\alpha < -2.5$. A separate component in the south-east is resolved and also detected at 610~MHz, showing a typical spectral index profile of a radio galaxy, but without a conclusive optical host for its AGN.

At 20\arcsec\, resolution significant diffuse emission is discovered throughout the cluster. We believe most of these areas in the outskirts of the cluster are a result of radio lobes expanding into less dense regions of the ICM, or from remnant plasma left over as radio galaxies move through the cluster. The area between the centre of mass of the two sub-clusters is bridged by steep-spectrum ($\alpha < -1.1$ given the non-detection in archival 610~MHz GMRT observations) low-surface brightness emission over hundreds of kpc resembling that of a radio halo. This strongly indicates that re-acceleration mechanisms are in play as a result of the cluster merger event. The presence of AGN throughout the cluster present a compelling mechanism for filling the ICM with relativistic electrons capable of fuelling the radio halo emission after the cluster merger event.

Five new giant radio galaxies are observed in the field of Abell 1682 where we find optical hosts with either photometric or spectroscopic redshifts in SDSS (that are not cluster members). They have physical sizes and luminosities (at 150~MHz) of: $2.86 \pm 0.55 \,\mathrm{Mpc}$, $1.56 \substack{+0.69 \\ -0.53} \times 10^{26} \,\mathrm{W/Hz}$ (FR-II); $0.68 \pm 0.20 \,\mathrm{Mpc}$, $4.84 \substack{+1.91 \\ -1.50} \times 10^{25} \,\mathrm{W/Hz}$ (FR-II); $2.07 \pm 0.18 \,\mathrm{Mpc}$, $3.04 \pm 0.46 \times 10^{26} \,\mathrm{W/Hz}$ (FR-I); $1.27 \pm 0.22 \,\mathrm{Mpc}$, $24.2 \pm 3.6 \,\mathrm{mJy}$ (FR-I); $1.08 \pm 0.36 \,\mathrm{Mpc}$, $2.33 \substack{+2.28 \\ -1.34} \times 10^{25} \,\mathrm{W/Hz}$ (FR-II). Furthermore, we find a giant remnant radio galaxy on the outskirts of Abell 1682 (associated with a cluster member) with a physical size of $1.29 \pm 0.29 \,\mathrm{Mpc}$, and luminosity of $9.61 \substack{+6.49 \\ -2.40} \times 10^{24} \,\mathrm{W/Hz}$.
 
\section*{acknowledgements}
We thank the referee for proving valuable comments helping to improve the clarity and interpretation of the science in this paper. A.O.C gives thanks to Ben Stappers for providing extensive comments on this paper as part of his PhD thesis. A.O.C, A.M.M.S and T.M.C gratefully acknowledge support from the European Research Council under grant ERC-2012-StG-307215 LODESTONE. A.B gratefully acknowledge support from the European Research Council under grant ERC-StG-714245 DRANOEL, n. 714245, and from the MIUR grant FARE SMS. RJvW acknowledges support from the ERC Advanced Investigator programme NewClusters 321271 and the VIDI research programme with project number 639.042.729, which is financed by the Netherlands Organisation for Scientific Research (NWO).

LOFAR, the Low Frequency Array designed and constructed by ASTRON, has facilities in several countries, that are owned by various parties (each with their own funding sources), and that are collectively operated by the International LOFAR Telescope (ILT) foundation under a joint scientific policy. 

This research has made use of the NASA/IPAC Extragalactic Database (NED) which is operated by the Jet Propulsion Laboratory, California Institute of Technology, under contract with the National Aeronautics and Space Administration.

This research made use of Numpy \citep{numpy}, APLpy \citep[an open-source plotting package for Python;][]{aplpy2012}, Matplotlib \citep{hunter2007}, and the cosmology calculator published in \cite{Wright_2006}.

Funding for the Sloan Digital Sky Survey IV has been provided by
the Alfred P. Sloan Foundation, the U.S. Department of Energy Office of
Science, and the Participating Institutions. SDSS-IV acknowledges
support and resources from the Center for High-Performance Computing at
the University of Utah. The SDSS web site is www.sdss.org.

SDSS-IV is managed by the Astrophysical Research Consortium for the
Participating Institutions of the SDSS Collaboration including the
Brazilian Participation Group, the Carnegie Institution for Science,
Carnegie Mellon University, the Chilean Participation Group, the French Participation Group, Harvard-Smithsonian Center for Astrophysics,
Instituto de Astrof\'isica de Canarias, The Johns Hopkins University,
Kavli Institute for the Physics and Mathematics of the Universe (IPMU) /
University of Tokyo, Lawrence Berkeley National Laboratory,
Leibniz Institut f\"ur Astrophysik Potsdam (AIP),
Max-Planck-Institut f\"ur Astronomie (MPIA Heidelberg),
Max-Planck-Institut f\"ur Astrophysik (MPA Garching),
Max-Planck-Institut f\"ur Extraterrestrische Physik (MPE),
National Astronomical Observatory of China, New Mexico State University,
New York University, University of Notre Dame,
Observat\'ario Nacional / MCTI, The Ohio State University,
Pennsylvania State University, Shanghai Astronomical Observatory,
United Kingdom Participation Group,
Universidad Nacional Aut\'onoma de M\'exico, University of Arizona,
University of Colorado Boulder, University of Oxford, University of Portsmouth,
University of Utah, University of Virginia, University of Washington, University of Wisconsin,
Vanderbilt University, and Yale University.


\bibliographystyle{aa}
\bibliography{thesis}

\begin{thebibliography}{68}
\expandafter\ifx\csname natexlab\endcsname\relax\def\natexlab#1{#1}\fi

\bibitem[{{Albareti} {et~al.}(2017){Albareti}, {Allende Prieto}, {Almeida},
  {Anders}, {Anderson}, {Andrews}, {Arag{\'o}n-Salamanca},
  {Argudo-Fern{\'a}ndez}, {Armengaud}, {Aubourg}, {Avila-Reese}, {Badenes},
  {Bailey}, {Barbuy}, {Barger}, {Barrera-Ballesteros}, {Bartosz}, {Basu},
  {Bates}, {Battaglia}, {Baumgarten}, {Baur}, {Bautista}, {Beers}, {Belfiore},
  {Bershady}, {Bertran de Lis}, {Bird}, {Bizyaev}, {Blanc}, {Blanton},
  {Blomqvist}, {Bolton}, {Borissova}, {Bovy}, {Brand t}, {Brinkmann},
  {Brownstein}, {Bundy}, {Burtin}, {Busca}, {Orlando Camacho Chavez}, {Cano
  D{\'\i}az}, {Cappellari}, {Carrera}, {Chen}, {Cherinka}, {Cheung},
  {Chiappini}, {Chojnowski}, {Chuang}, {Chung}, {Cirolini}, {Clerc}, {Cohen},
  {Comerford}, {Comparat}, {Correa do Nascimento}, {Cousinou}, {Covey},
  {Crane}, {Croft}, {Cunha}, {Darling}, {Davidson}, {Dawson}, {Da Costa}, {Da
  Silva Ilha}, {Deconto Machado}, {Delubac}, {De Lee}, {De la Macorra}, {De la
  Torre}, {Diamond-Stanic}, {Donor}, {Downes}, {Drory}, {Du}, {Du Mas des
  Bourboux}, {Dwelly}, {Ebelke}, {Eigenbrot}, {Eisenstein}, {Elsworth},
  {Emsellem}, {Eracleous}, {Escoffier}, {Evans}, {Falc{\'o}n-Barroso}, {Fan},
  {Favole}, {Fernandez-Alvar}, {Fernand ez-Trincado}, {Feuillet}, {Fleming},
  {Font-Ribera}, {Freischlad}, {Frinchaboy}, {Fu}, {Gao}, {Garcia},
  {Garcia-Dias}, {Garcia-Hern{\'a}ndez}, {Garcia P{\'e}rez}, {Gaulme}, {Ge},
  {Geisler}, {Gillespie}, {Gil Marin}, {Girardi}, {Goddard}, {Gomez Maqueo
  Chew}, {Gonzalez-Perez}, {Grabowski}, {Green}, {Grier}, {Grier}, {Guo},
  {Guy}, {Hagen}, {Hall}, {Harding}, {Harley}, {Hasselquist}, {Hawley},
  {Hayes}, {Hearty}, {Hekker}, {Hernandez Toledo}, {Ho}, {Hogg},
  {Holley-Bockelmann}, {Holtzman}, {Holzer}, {Hu}, {Huber}, {Hutchinson},
  {Hwang}, {Ibarra-Medel}, {Ivans}, {Ivory}, {Jaehnig}, {Jensen}, {Johnson},
  {Jones}, {Jullo}, {Kallinger}, {Kinemuchi}, {Kirkby}, {Klaene}, {Kneib},
  {Kollmeier}, {Lacerna}, {Lane}, {Lang}, {Laurent}, {Law}, {Leauthaud}, {Le
  Goff}, {Li}, {Li}, {Li}, {Li}, {Liang}, {Liang}, {Lima}, {Lin}, {Lin}, {Lin},
  {Liu}, {Long}, {Lucatello}, {MacDonald}, {MacLeod}, {Mackereth}, {Mahadevan},
  {Maia}, {Maiolino}, {Majewski}, {Malanushenko}, {Malanushenko}, {Mallmann},
  {Manchado}, {Maraston}, {Marques-Chaves}, {Martinez Valpuesta}, {Masters},
  {Mathur}, {McGreer}, {Merloni}, {Merrifield}, {M{\'e}sz{\'a}ros}, {Meza},
  {Miglio}, {Minchev}, {Molaverdikhani}, {Montero-Dorta}, {Mosser}, {Muna},
  {Myers}, {Nair}, {Nandra}, {Ness}, {Newman}, {Nichol}, {Nidever},
  {Nitschelm}, {O'Connell}, {Oravetz}, {Oravetz}, {Pace}, {Padilla},
  {Palanque-Delabrouille}, {Pan}, {Parejko}, {Paris}, {Park}, {Peacock},
  {Peirani}, {Pellejero-Ibanez}, {Penny}, {Percival}, {Percival},
  {Perez-Fournon}, {Petitjean}, {Pieri}, {Pinsonneault}, {Pisani}, {Prada},
  {Prakash}, {Price-Jones}, {Raddick}, {Rahman}, {Raichoor}, {Barboza Rembold},
  {Reyna}, {Rich}, {Richstein}, {Ridl}, {Riffel}, {Riffel}, {Rix}, {Robin},
  {Rockosi}, {Rodr{\'\i}guez-Torres}, {Rodrigues}, {Roe}, {Roman Lopes},
  {Rom{\'a}n-Z{\'u}{\~n}iga}, {Ross}, {Rossi}, {Ruan}, {Ruggeri}, {Runnoe},
  {Salazar-Albornoz}, {Salvato}, {Sanchez}, {Sanchez}, {Sanchez-Gallego},
  {Santiago}, {Schiavon}, {Schimoia}, {Schlafly}, {Schlegel}, {Schneider},
  {Sch{\"o}nrich}, {Schultheis}, {Schwope}, {Seo}, {Serenelli}, {Sesar},
  {Shao}, {Shetrone}, {Shull}, {Silva Aguirre}, {Skrutskie}, {Slosar}, {Smith},
  {Smith}, {Sobeck}, {Somers}, {Souto}, {Stark}, {Stassun}, {Steinmetz},
  {Stello}, {Storchi Bergmann}, {Strauss}, {Streblyanska}, {Stringfellow},
  {Suarez}, {Sun}, {Taghizadeh-Popp}, {Tang}, {Tao}, {Tayar}, {Tembe},
  {Thomas}, {Tinker}, {Tojeiro}, {Tremonti}, {Troup}, {Trump}, {Unda-Sanzana},
  {Valenzuela}, {Van den Bosch}, {Vargas-Maga{\~n}a}, {Vazquez}, {Villanova},
  {Vivek}, {Vogt}, {Wake}, {Walterbos}, {Wang}, {Wang}, {Weaver}, {Weijmans},
  {Weinberg}, {Westfall}, {Whelan}, {Wilcots}, {Wild}, {Williams}, {Wilson},
  {Wood-Vasey}, {Wylezalek}, {Xiao}, {Yan}, {Yang}, {Ybarra}, {Yeche}, {Yuan},
  {Zakamska}, {Zamora}, {Zasowski}, {Zhang}, {Zhao}, {Zhao}, {Zheng}, {Zheng},
  {Zhou}, {Zhu}, {Zinn}, \& {Zou}}]{SDSSDR132017}
{Albareti}, F.~D., {Allende Prieto}, C., {Almeida}, A., {et~al.} 2017, The
  Astrophysical Journal Supplement Series, 233, 25

\bibitem[{{Allen} {et~al.}(1992){Allen}, {Edge}, {Fabian}, {Bohringer},
  {Crawford}, {Ebeling}, {Johnstone}, {Naylor}, \& {Schwarz}}]{allen1992}
{Allen}, S.~W., {Edge}, A.~C., {Fabian}, A.~C., {et~al.} 1992, \mnras, 259, 67

\bibitem[{{Beck} {et~al.}(2016){Beck}, {Dobos}, {Budav{\'a}ri}, {Szalay}, \&
  {Csabai}}]{beck2016}
{Beck}, R., {Dobos}, L., {Budav{\'a}ri}, T., {Szalay}, A.~S., \& {Csabai}, I.
  2016, \mnras, 460, 1371

\bibitem[{{Blandford} \& {Eichler}(1987)}]{blandford1987}
{Blandford}, R. \& {Eichler}, D. 1987, \physrep, 154, 1

\bibitem[{{Botteon} {et~al.}(2016){Botteon}, {Gastaldello}, {Brunetti}, \&
  {Dallacasa}}]{botteon2016}
{Botteon}, A., {Gastaldello}, F., {Brunetti}, G., \& {Dallacasa}, D. 2016,
  \mnras, 460, L84

\bibitem[{{Botteon} {et~al.}(2018){Botteon}, {Shimwell}, {Bonafede},
  {Dallacasa}, {Brunetti}, {Mandal}, {van Weeren}, {Br{\"u}ggen}, {Cassano},
  {de Gasperin}, {Hoang}, {Hoeft}, {R{\"o}ttgering}, {Savini}, {White},
  {Wilber}, \& {Venturi}}]{Botteon2018}
{Botteon}, A., {Shimwell}, T.~W., {Bonafede}, A., {et~al.} 2018, \mnras, 478,
  885

\bibitem[{{Brienza} {et~al.}(2017){Brienza}, {Godfrey}, {Morganti}, {Prandoni},
  {Harwood}, {Mahony}, {Hardcastle}, {Murgia}, {R{\"o}ttgering}, {Shimwell}, \&
  {Shulevski}}]{brienza2017}
{Brienza}, M., {Godfrey}, L., {Morganti}, R., {et~al.} 2017, \aap, 606, A98

\bibitem[{{Briggs}(1995)}]{Briggs}
{Briggs}, D., S. 1995, New Mexico Institute of Mining Technology, Socorro, New
  Mexico, USA

\bibitem[{{Brunetti} {et~al.}(2008){Brunetti}, {Giacintucci}, {Cassano},
  {Lane}, {Dallacasa}, {Venturi}, {Kassim}, {Setti}, {Cotton}, \&
  {Markevitch}}]{brunetti2008nat}
{Brunetti}, G., {Giacintucci}, S., {Cassano}, R., {et~al.} 2008, \nat, 455, 944

\bibitem[{{Brunetti} \& {Jones}(2014)}]{brunetti2014}
{Brunetti}, G. \& {Jones}, T.~W. 2014, International Journal of Modern Physics
  D, 23, 1430007

\bibitem[{{Brunetti} \& {Lazarian}(2007)}]{BrunettiLazarian2007}
{Brunetti}, G. \& {Lazarian}, A. 2007, \mnras, 378, 245

\bibitem[{{Brunetti} \& {Lazarian}(2011)}]{BrunettiLazarian2011}
{Brunetti}, G. \& {Lazarian}, A. 2011, \mnras, 410, 127

\bibitem[{{Brunetti} {et~al.}(2001){Brunetti}, {Setti}, {Feretti}, \&
  {Giovannini}}]{brunetti2001}
{Brunetti}, G., {Setti}, G., {Feretti}, L., \& {Giovannini}, G. 2001, \mnras,
  320, 365

\bibitem[{{Cassano} {et~al.}(2006){Cassano}, {Brunetti}, \&
  {Setti}}]{cassano2006}
{Cassano}, R., {Brunetti}, G., \& {Setti}, G. 2006, \mnras, 369, 1577

\bibitem[{{Dahle} {et~al.}(2002){Dahle}, {Kaiser}, {Irgens}, {Lilje}, \&
  {Maddox}}]{dahle2002}
{Dahle}, H., {Kaiser}, N., {Irgens}, R.~J., {Lilje}, P.~B., \& {Maddox}, S.~J.
  2002, \apjs, 139, 313

\bibitem[{{de Gasperin}(2017)}]{gasperin2017}
{de Gasperin}, F. 2017, \mnras, 467, 2234

\bibitem[{{de Gasperin} {et~al.}(2017){de Gasperin}, {Intema}, {Shimwell},
  {Brunetti}, {Br{\"u}ggen}, {En{\ss}lin}, {van Weeren}, {Bonafede}, \&
  {R{\"o}ttgering}}]{gasperinGREET2017}
{de Gasperin}, F., {Intema}, H.~T., {Shimwell}, T.~W., {et~al.} 2017, Science
  Advances, 3, e1701634

\bibitem[{{Ebeling} {et~al.}(1998){Ebeling}, {Edge}, {Bohringer}, {Allen},
  {Crawford}, {Fabian}, {Voges}, \& {Huchra}}]{ebeling1998}
{Ebeling}, H., {Edge}, A.~C., {Bohringer}, H., {et~al.} 1998, \mnras, 301, 881

\bibitem[{{Eckert} {et~al.}(2016){Eckert}, {Jauzac}, {Vazza}, {Owers}, {Kneib},
  {Tchernin}, {Intema}, \& {Knowles}}]{eckert2016}
{Eckert}, D., {Jauzac}, M., {Vazza}, F., {et~al.} 2016, \mnras, 461, 1302

\bibitem[{{Eisenstein} {et~al.}(2011){Eisenstein}, {Weinberg}, {Agol},
  {Aihara}, {Allende Prieto}, {Anderson}, {Arns}, {Aubourg}, {Bailey},
  {Balbinot}, \& et~al.}]{SDSS}
{Eisenstein}, D.~J., {Weinberg}, D.~H., {Agol}, E., {et~al.} 2011, \apj, 142,
  72

\bibitem[{{Ensslin} {et~al.}(1998){Ensslin}, {Biermann}, {Klein}, \&
  {Kohle}}]{ensslin1998}
{Ensslin}, T.~A., {Biermann}, P.~L., {Klein}, U., \& {Kohle}, S. 1998, \aap,
  332, 395

\bibitem[{{En{\ss}lin} \& {Gopal-Krishna}(2001)}]{ensslin2001}
{En{\ss}lin}, T.~A. \& {Gopal-Krishna}. 2001, \aap, 366, 26

\bibitem[{{Fanaroff} \& {Riley}(1974)}]{FR1974}
{Fanaroff}, B.~L. \& {Riley}, J.~M. 1974, \mnras, 167, 31P

\bibitem[{{Flewelling} {et~al.}(2016){Flewelling}, {Magnier}, {Chambers},
  {Heasley}, {Holmberg}, {Huber}, {Sweeney}, {Waters}, {Calamida}, {Casertano},
  {Chen}, {Farrow}, {Hasinger}, {Henderson}, {Long}, {Metcalfe}, {Narayan},
  {Nieto-Santisteban}, {Norberg}, {Rest}, {Saglia}, {Szalay}, {Thakar},
  {Tonry}, {Valenti}, {Werner}, {White}, {Denneau}, {Draper}, {Hodapp},
  {Jedicke}, {Kaiser}, {Kudritzki}, {Price}, {Wainscoat}, {Builders},
  {Chastel}, {McLean}, {Postman}, \& {Shiao}}]{panstarrs2016}
{Flewelling}, H.~A., {Magnier}, E.~A., {Chambers}, K.~C., {et~al.} 2016, arXiv
  e-prints, arXiv:1612.05243

\bibitem[{{Guo} {et~al.}(2014{\natexlab{a}}){Guo}, {Sironi}, \&
  {Narayan}}]{guo2014a}
{Guo}, X., {Sironi}, L., \& {Narayan}, R. 2014{\natexlab{a}}, \apj, 794, 153

\bibitem[{{Guo} {et~al.}(2014{\natexlab{b}}){Guo}, {Sironi}, \&
  {Narayan}}]{guo2014b}
{Guo}, X., {Sironi}, L., \& {Narayan}, R. 2014{\natexlab{b}}, \apj, 797, 47

\bibitem[{{Harwood} {et~al.}(2015){Harwood}, {Hardcastle}, \&
  {Croston}}]{harwood2015}
{Harwood}, J.~J., {Hardcastle}, M.~J., \& {Croston}, J.~H. 2015, \mnras, 454,
  3403

\bibitem[{{Harwood} {et~al.}(2013){Harwood}, {Hardcastle}, {Croston}, \&
  {Goodger}}]{harwood2013}
{Harwood}, J.~J., {Hardcastle}, M.~J., {Croston}, J.~H., \& {Goodger}, J.~L.
  2013, \mnras, 435, 3353

\bibitem[{{Hoang} {et~al.}(2017){Hoang}, {Shimwell}, {Stroe}, {Akamatsu},
  {Brunetti}, {Donnert}, {Intema}, {Mulcahy}, {R{\"o}ttgering}, {van Weeren},
  {Bonafede}, {Br{\"u}ggen}, {Cassano}, {Chy{\.z}y}, {En{\ss}lin}, {Ferrari},
  {de Gasperin}, {Gu}, {Hoeft}, {Miley}, {Orr{\'u}}, {Pizzo}, \&
  {White}}]{hoang2017}
{Hoang}, D.~N., {Shimwell}, T.~W., {Stroe}, A., {et~al.} 2017, \mnras, 471,
  1107

\bibitem[{Hunter(2007)}]{hunter2007}
Hunter, J.~D. 2007, Computing In Science \& Engineering, 9, 90

\bibitem[{{Intema}(2014)}]{intema2014}
{Intema}, H.~T. 2014, in Astronomical Society of India Conference Series,
  Vol.~13, Astronomical Society of India Conference Series, 469

\bibitem[{{Intema} {et~al.}(2017){Intema}, {Jagannathan}, {Mooley}, \&
  {Frail}}]{intema2017}
{Intema}, H.~T., {Jagannathan}, P., {Mooley}, K.~P., \& {Frail}, D.~A. 2017,
  \aap, 598, A78

\bibitem[{{Lane} {et~al.}(2014){Lane}, {Cotton}, {van Velzen}, {Clarke},
  {Kassim}, {Helmboldt}, {Lazio}, \& {Cohen}}]{VLSSr2014}
{Lane}, W.~M., {Cotton}, W.~D., {van Velzen}, S., {et~al.} 2014, \mnras, 440,
  327

\bibitem[{{Macario} {et~al.}(2013){Macario}, {Venturi}, {Intema}, {Dallacasa},
  {Brunetti}, {Cassano}, {Giacintucci}, {Ferrari}, {Ishwara-Chandra}, \&
  {Athreya}}]{macario2013}
{Macario}, G., {Venturi}, T., {Intema}, H.~T., {et~al.} 2013, \aap, 551, A141

\bibitem[{{Mohan} \& {Rafferty}(2015)}]{pybdsf2015}
{Mohan}, N. \& {Rafferty}, D. 2015, {PyBDSF: Python Blob Detection and Source
  Finder}

\bibitem[{{Murgia} {et~al.}(2011){Murgia}, {Parma}, {Mack}, {de Ruiter},
  {Fanti}, {Govoni}, {Tarchi}, {Giacintucci}, \& {Markevitch}}]{Murgia2011}
{Murgia}, M., {Parma}, P., {Mack}, K.~H., {et~al.} 2011, \aap, 526, A148

\bibitem[{{Nuza} {et~al.}(2017){Nuza}, {Gelszinnis}, {Hoeft}, \&
  {Yepes}}]{nuza2017}
{Nuza}, S.~E., {Gelszinnis}, J., {Hoeft}, M., \& {Yepes}, G. 2017, \mnras, 470,
  240

\bibitem[{Oliphant(2006--)}]{numpy}
Oliphant, T. 2006--, {NumPy}: A guide to {NumPy}, USA: Trelgol Publishing,
  [Online; accessed <today>]

\bibitem[{{Pandey} {et~al.}(2009){Pandey}, {van Zwieten}, {de Bruyn}, \&
  {Nijboer}}]{Pandey2009}
{Pandey}, V.~N., {van Zwieten}, J.~E., {de Bruyn}, A.~G., \& {Nijboer}, R.
  2009, in Astronomical Society of the Pacific Conference Series, Vol. 407, The
  Low-Frequency Radio Universe, ed. D.~J. {Saikia}, D.~A. {Green}, Y.~{Gupta},
  \& T.~{Venturi}, 384

\bibitem[{{Petrosian}(2001)}]{petrosian2001}
{Petrosian}, V. 2001, \apj, 557, 560

\bibitem[{{Pinzke} {et~al.}(2015){Pinzke}, {Oh}, \& {Pfrommer}}]{pinzke2015}
{Pinzke}, A., {Oh}, S.~P., \& {Pfrommer}, C. 2015, ArXiv e-prints

\bibitem[{{Pinzke} {et~al.}(2017){Pinzke}, {Oh}, \& {Pfrommer}}]{pinzke2017}
{Pinzke}, A., {Oh}, S.~P., \& {Pfrommer}, C. 2017, \mnras, 465, 4800

\bibitem[{{Planck Collaboration} {et~al.}(2015){Planck Collaboration}, {Ade},
  {Aghanim}, {Armitage-Caplan}, {Arnaud}, {Ashdown}, {Atrio-Barandela},
  {Aumont}, {Aussel}, {Baccigalupi}, \& et~al.}]{plankmass2013}
{Planck Collaboration}, {Ade}, P.~A.~R., {Aghanim}, N., {et~al.} 2015, \aap,
  581, A14

\bibitem[{{Planck Collaboration} {et~al.}(2016{\natexlab{a}}){Planck
  Collaboration}, {Ade}, {Aghanim}, {Arnaud}, {Ashdown}, {Aumont},
  {Baccigalupi}, {Banday}, {Barreiro}, {Barrena}, \& et~al.}]{plankmass2015}
{Planck Collaboration}, {Ade}, P.~A.~R., {Aghanim}, N., {et~al.}
  2016{\natexlab{a}}, \aap, 594, A27

\bibitem[{{Planck Collaboration} {et~al.}(2016{\natexlab{b}}){Planck
  Collaboration}, {Ade}, {Aghanim}, {Arnaud}, {Ashdown}, {Aumont},
  {Baccigalupi}, {Banday}, {Barreiro}, {Bartlett}, {Bartolo}, {Battaner},
  {Battye}, {Benabed}, {Beno{\^\i}t}, {Benoit-L{\'e}vy}, {Bernard},
  {Bersanelli}, {Bielewicz}, {Bock}, {Bonaldi}, {Bonavera}, {Bond}, {Borrill},
  {Bouchet}, {Boulanger}, {Bucher}, {Burigana}, {Butler}, {Calabrese},
  {Cardoso}, {Catalano}, {Challinor}, {Chamballu}, {Chary}, {Chiang}, {Chluba},
  {Christensen}, {Church}, {Clements}, {Colombi}, {Colombo}, {Combet},
  {Coulais}, {Crill}, {Curto}, {Cuttaia}, {Danese}, {Davies}, {Davis}, {de
  Bernardis}, {de Rosa}, {de Zotti}, {Delabrouille}, {D{\'e}sert}, {Di
  Valentino}, {Dickinson}, {Diego}, {Dolag}, {Dole}, {Donzelli}, {Dor{\'e}},
  {Douspis}, {Ducout}, {Dunkley}, {Dupac}, {Efstathiou}, {Elsner},
  {En{\ss}lin}, {Eriksen}, {Farhang}, {Fergusson}, {Finelli}, {Forni},
  {Frailis}, {Fraisse}, {Franceschi}, {Frejsel}, {Galeotta}, {Galli}, {Ganga},
  {Gauthier}, {Gerbino}, {Ghosh}, {Giard}, {Giraud-H{\'e}raud}, {Giusarma},
  {Gjerl{\o}w}, {Gonz{\'a}lez-Nuevo}, {G{\'o}rski}, {Gratton}, {Gregorio},
  {Gruppuso}, {Gudmundsson}, {Hamann}, {Hansen}, {Hanson}, {Harrison}, {Helou},
  {Henrot-Versill{\'e}}, {Hern{\'a}ndez-Monteagudo}, {Herranz}, {Hildebrand t},
  {Hivon}, {Hobson}, {Holmes}, {Hornstrup}, {Hovest}, {Huang}, {Huffenberger},
  {Hurier}, {Jaffe}, {Jaffe}, {Jones}, {Juvela}, {Keih{\"a}nen}, {Keskitalo},
  {Kisner}, {Kneissl}, {Knoche}, {Knox}, {Kunz}, {Kurki-Suonio}, {Lagache},
  {L{\"a}hteenm{\"a}ki}, {Lamarre}, {Lasenby}, {Lattanzi}, {Lawrence}, {Leahy},
  {Leonardi}, {Lesgourgues}, {Levrier}, {Lewis}, {Liguori}, {Lilje},
  {Linden-V{\o}rnle}, {L{\'o}pez-Caniego}, {Lubin}, {Mac{\'\i}as-P{\'e}rez},
  {Maggio}, {Maino}, {Mandolesi}, {Mangilli}, {Marchini}, {Maris}, {Martin},
  {Martinelli}, {Mart{\'\i}nez-Gonz{\'a}lez}, {Masi}, {Matarrese}, {McGehee},
  {Meinhold}, {Melchiorri}, {Melin}, {Mendes}, {Mennella}, {Migliaccio},
  {Millea}, {Mitra}, {Miville-Desch{\^e}nes}, {Moneti}, {Montier}, {Morgante},
  {Mortlock}, {Moss}, {Munshi}, {Murphy}, {Naselsky}, {Nati}, {Natoli},
  {Netterfield}, {N{\o}rgaard-Nielsen}, {Noviello}, {Novikov}, {Novikov},
  {Oxborrow}, {Paci}, {Pagano}, {Pajot}, {Paladini}, {Paoletti}, {Partridge},
  {Pasian}, {Patanchon}, {Pearson}, {Perdereau}, {Perotto}, {Perrotta},
  {Pettorino}, {Piacentini}, {Piat}, {Pierpaoli}, {Pietrobon}, {Plaszczynski},
  {Pointecouteau}, {Polenta}, {Popa}, {Pratt}, {Pr{\'e}zeau}, {Prunet},
  {Puget}, {Rachen}, {Reach}, {Rebolo}, {Reinecke}, {Remazeilles}, {Renault},
  {Renzi}, {Ristorcelli}, {Rocha}, {Rosset}, {Rossetti}, {Roudier},
  {Rouill{\'e} d'Orfeuil}, {Rowan-Robinson}, {Rubi{\~n}o-Mart{\'\i}n},
  {Rusholme}, {Said}, {Salvatelli}, {Salvati}, {Sandri}, {Santos},
  {Savelainen}, {Savini}, {Scott}, {Seiffert}, {Serra}, {Shellard}, {Spencer},
  {Spinelli}, {Stolyarov}, {Stompor}, {Sudiwala}, {Sunyaev}, {Sutton},
  {Suur-Uski}, {Sygnet}, {Tauber}, {Terenzi}, {Toffolatti}, {Tomasi},
  {Tristram}, {Trombetti}, {Tucci}, {Tuovinen}, {T{\"u}rler}, {Umana},
  {Valenziano}, {Valiviita}, {Van Tent}, {Vielva}, {Villa}, {Wade}, {Wandelt},
  {Wehus}, {White}, {White}, {Wilkinson}, {Yvon}, {Zacchei}, \&
  {Zonca}}]{plank2016}
{Planck Collaboration}, {Ade}, P.~A.~R., {Aghanim}, N., {et~al.}
  2016{\natexlab{b}}, \aap, 594, A13

\bibitem[{{Robitaille} \& {Bressert}(2012)}]{aplpy2012}
{Robitaille}, T. \& {Bressert}, E. 2012, {APLpy: Astronomical Plotting Library
  in Python}, Astrophysics Source Code Library

\bibitem[{{Sakelliou} \& {Merrifield}(2000)}]{sakelliou2000}
{Sakelliou}, I. \& {Merrifield}, M.~R. 2000, \mnras, 311, 649

\bibitem[{{Scaife} \& {Heald}(2012)}]{scaifeheald2012}
{Scaife}, A.~M.~M. \& {Heald}, G.~H. 2012, \mnras, 423, L30

\bibitem[{{Shimwell} {et~al.}(2016){Shimwell}, {Luckin}, {Br{\"u}ggen},
  {Brunetti}, {Intema}, {Owers}, {R{\"o}ttgering}, {Stroe}, {van Weeren},
  {Williams}, {Cassano}, {de Gasperin}, {Heald}, {Hoang}, {Hardcastle},
  {Sridhar}, {Sabater}, {Best}, {Bonafede}, {Chy{\.z}y}, {En{\ss}lin},
  {Ferrari}, {Haverkorn}, {Hoeft}, {Horellou}, {McKean}, {Morabito},
  {Orr{\`u}}, {Pizzo}, {Retana-Montenegro}, \& {White}}]{shimwell2016}
{Shimwell}, T.~W., {Luckin}, J., {Br{\"u}ggen}, M., {et~al.} 2016, \mnras, 459,
  277

\bibitem[{{Shimwell} {et~al.}(2019){Shimwell}, {Tasse}, {Hardcastle}, {Mechev},
  {Williams}, {Best}, {R{\"o}ttgering}, {Callingham}, {Dijkema}, {de Gasperin},
  {Hoang}, {Hugo}, {Mirmont}, {Oonk}, {Prandoni}, {Rafferty}, {Sabater},
  {Smirnov}, {van Weeren}, {White}, {Atemkeng}, {Bester}, {Bonnassieux},
  {Br{\"u}ggen}, {Brunetti}, {Chy{\.z}y}, {Cochrane}, {Conway}, {Croston},
  {Danezi}, {Duncan}, {Haverkorn}, {Heald}, {Iacobelli}, {Intema}, {Jackson},
  {Jamrozy}, {Jarvis}, {Lakhoo}, {Mevius}, {Miley}, {Morabito}, {Morganti},
  {Nisbet}, {Orr{\'u}}, {Perkins}, {Pizzo}, {Schrijvers}, {Smith}, {Vermeulen},
  {Wise}, {Alegre}, {Bacon}, {van Bemmel}, {Beswick}, {Bonafede}, {Botteon},
  {Bourke}, {Brienza}, {Calistro Rivera}, {Cassano}, {Clarke}, {Conselice},
  {Dettmar}, {Drabent}, {Dumba}, {Emig}, {En{\ss}lin}, {Ferrari}, {Garrett},
  {G{\'e}nova-Santos}, {Goyal}, {G{\"u}rkan}, {Hale}, {Harwood}, {Heesen},
  {Hoeft}, {Horellou}, {Jackson}, {Kokotanekov}, {Kondapally},
  {Kunert-Bajraszewska}, {Mahatma}, {Mahony}, {Mandal}, {McKean}, {Merloni},
  {Mingo}, {Miskolczi}, {Mooney}, {Nikiel-Wroczy{\'n}ski}, {O'Sullivan},
  {Quinn}, {Reich}, {Roskowi{\'n}ski}, {Rowlinson}, {Savini}, {Saxena},
  {Schwarz}, {Shulevski}, {Sridhar}, {Stacey}, {Urquhart}, {van der Wiel},
  {Varenius}, {Webster}, \& {Wilber}}]{shimwell2019}
{Shimwell}, T.~W., {Tasse}, C., {Hardcastle}, M.~J., {et~al.} 2019, \aap, 622,
  A1

\bibitem[{{Smirnov} \& {Tasse}(2015)}]{Smirnov2015}
{Smirnov}, O.~M. \& {Tasse}, C. 2015, \mnras, 449, 2668

\bibitem[{{Stocke} \& {Burns}(1987)}]{stocke1987}
{Stocke}, J.~T. \& {Burns}, J.~O. 1987, \apj, 319, 671

\bibitem[{{Stroe} {et~al.}(2014){Stroe}, {Harwood}, {Hardcastle}, \&
  {R{\"o}ttgering}}]{stroe2014}
{Stroe}, A., {Harwood}, J.~J., {Hardcastle}, M.~J., \& {R{\"o}ttgering},
  H.~J.~A. 2014, \mnras, 445, 1213

\bibitem[{{Swarup}(1991)}]{GMRT}
{Swarup}, G. 1991, in Astronomical Society of the Pacific Conference Series,
  Vol.~19, IAU Colloq. 131: Radio Interferometry. Theory, Techniques, and
  Applications, ed. T.~J. {Cornwell} \& R.~A. {Perley}, 376--380

\bibitem[{{Tasse}(2014)}]{Tasse2014}
{Tasse}, C. 2014, arXiv e-prints, arXiv:1410.8706

\bibitem[{{Tasse} {et~al.}(2018){Tasse}, {Hugo}, {Mirmont}, {Smirnov},
  {Atemkeng}, {Bester}, {Hardcastle}, {Lakhoo}, {Perkins}, \&
  {Shimwell}}]{Tasse2018}
{Tasse}, C., {Hugo}, B., {Mirmont}, M., {et~al.} 2018, \aap, 611, A87

\bibitem[{{van Diepen} \& {Dijkema}(2018)}]{Diepen2018}
{van Diepen}, G. \& {Dijkema}, T.~J. 2018, {DPPP: Default Pre-Processing
  Pipeline}

\bibitem[{{van Haarlem} {et~al.}(2013){van Haarlem}, {Wise}, {Gunst}, {Heald},
  {McKean}, {Hessels}, {de Bruyn}, {Nijboer}, {Swinbank}, {Fallows},
  {Brentjens}, {Nelles}, {Beck}, {Falcke}, {Fender}, {H{\"o}randel},
  {Koopmans}, {Mann}, {Miley}, {R{\"o}ttgering}, {Stappers}, {Wijers},
  {Zaroubi}, {van den Akker}, {Alexov}, {Anderson}, {Anderson}, {van Ardenne},
  {Arts}, {Asgekar}, {Avruch}, {Batejat}, {B{\"a}hren}, {Bell}, {Bell}, {van
  Bemmel}, {Bennema}, {Bentum}, {Bernardi}, {Best}, {B{\^i}rzan}, {Bonafede},
  {Boonstra}, {Braun}, {Bregman}, {Breitling}, {van de Brink}, {Broderick},
  {Broekema}, {Brouw}, {Br{\"u}ggen}, {Butcher}, {van Cappellen}, {Ciardi},
  {Coenen}, {Conway}, {Coolen}, {Corstanje}, {Damstra}, {Davies}, {Deller},
  {Dettmar}, {van Diepen}, {Dijkstra}, {Donker}, {Doorduin}, {Dromer}, {Drost},
  {van Duin}, {Eisl{\"o}ffel}, {van Enst}, {Ferrari}, {Frieswijk}, {Gankema},
  {Garrett}, {de Gasperin}, {Gerbers}, {de Geus}, {Grie{\ss}meier}, {Grit},
  {Gruppen}, {Hamaker}, {Hassall}, {Hoeft}, {Holties}, {Horneffer}, {van der
  Horst}, {van Houwelingen}, {Huijgen}, {Iacobelli}, {Intema}, {Jackson},
  {Jelic}, {de Jong}, {Juette}, {Kant}, {Karastergiou}, {Koers}, {Kollen},
  {Kondratiev}, {Kooistra}, {Koopman}, {Koster}, {Kuniyoshi}, {Kramer},
  {Kuper}, {Lambropoulos}, {Law}, {van Leeuwen}, {Lemaitre}, {Loose}, {Maat},
  {Macario}, {Markoff}, {Masters}, {McFadden}, {McKay-Bukowski}, {Meijering},
  {Meulman}, {Mevius}, {Middelberg}, {Millenaar}, {Miller-Jones}, {Mohan},
  {Mol}, {Morawietz}, {Morganti}, {Mulcahy}, {Mulder}, {Munk}, {Nieuwenhuis},
  {van Nieuwpoort}, {Noordam}, {Norden}, {Noutsos}, {Offringa}, {Olofsson},
  {Omar}, {Orr{\'u}}, {Overeem}, {Paas}, {Pandey-Pommier}, {Pandey}, {Pizzo},
  {Polatidis}, {Rafferty}, {Rawlings}, {Reich}, {de Reijer}, {Reitsma},
  {Renting}, {Riemers}, {Rol}, {Romein}, {Roosjen}, {Ruiter}, {Scaife}, {van
  der Schaaf}, {Scheers}, {Schellart}, {Schoenmakers}, {Schoonderbeek},
  {Serylak}, {Shulevski}, {Sluman}, {Smirnov}, {Sobey}, {Spreeuw}, {Steinmetz},
  {Sterks}, {Stiepel}, {Stuurwold}, {Tagger}, {Tang}, {Tasse}, {Thomas},
  {Thoudam}, {Toribio}, {van der Tol}, {Usov}, {van Veelen}, {van der Veen},
  {ter Veen}, {Verbiest}, {Vermeulen}, {Vermaas}, {Vocks}, {Vogt}, {de Vos},
  {van der Wal}, {van Weeren}, {Weggemans}, {Weltevrede}, {White}, {Wijnholds},
  {Wilhelmsson}, {Wucknitz}, {Yatawatta}, {Zarka}, {Zensus}, \& {van
  Zwieten}}]{LOFAR}
{van Haarlem}, M.~P., {Wise}, M.~W., {Gunst}, A.~W., {et~al.} 2013, \aap, 556,
  A2

\bibitem[{{van Weeren} {et~al.}(2017){van Weeren}, {Andrade-Santos}, {Dawson},
  {Golovich}, {Lal}, {Kang}, {Ryu}, {Br{\`i}ggen}, {Ogrean}, {Forman}, {Jones},
  {Placco}, {Santucci}, {Wittman}, {Jee}, {Kraft}, {Sobral}, {Stroe}, \&
  {Fogarty}}]{weeren2017}
{van Weeren}, R.~J., {Andrade-Santos}, F., {Dawson}, W.~A., {et~al.} 2017,
  Nature Astronomy, 1, 0005

\bibitem[{{van Weeren} {et~al.}(2016{\natexlab{a}}){van Weeren}, {Brunetti},
  {Br{\"u}ggen}, {Andrade-Santos}, {Ogrean}, {Williams}, {R{\"o}ttgering},
  {Dawson}, {Forman}, {de Gasperin}, {Hardcastle}, {Jones}, {Miley},
  {Rafferty}, {Rudnick}, {Sabater}, {Sarazin}, {Shimwell}, {Bonafede}, {Best},
  {B{\^i}rzan}, {Cassano}, {Chy{\.z}y}, {Croston}, {Dijkema}, {En{\ss}lin},
  {Ferrari}, {Heald}, {Hoeft}, {Horellou}, {Jarvis}, {Kraft}, {Mevius},
  {Intema}, {Murray}, {Orr{\'u}}, {Pizzo}, {Sridhar}, {Simionescu}, {Stroe},
  {van der Tol}, \& {White}}]{weeren2016toothbrush}
{van Weeren}, R.~J., {Brunetti}, G., {Br{\"u}ggen}, M., {et~al.}
  2016{\natexlab{a}}, \apj, 818, 204

\bibitem[{{van Weeren} {et~al.}(2019){van Weeren}, {de Gasperin}, {Akamatsu},
  {Br{\"u}ggen}, {Feretti}, {Kang}, {Stroe}, \& {Zandanel}}]{weerenreview2019}
{van Weeren}, R.~J., {de Gasperin}, F., {Akamatsu}, H., {et~al.} 2019, \ssr,
  215, 16

\bibitem[{{van Weeren} {et~al.}(2016{\natexlab{b}}){van Weeren}, {Williams},
  {Hardcastle}, {Shimwell}, {Rafferty}, {Sabater}, {Heald}, {Sridhar},
  {Dijkema}, {Brunetti}, {Br{\"u}ggen}, {Andrade-Santos}, {Ogrean},
  {R{\"o}ttgering}, {Dawson}, {Forman}, {de Gasperin}, {Jones}, {Miley},
  {Rudnick}, {Sarazin}, {Bonafede}, {Best}, {B{\^i}rzan}, {Cassano},
  {Chy{\.z}y}, {Croston}, {Ensslin}, {Ferrari}, {Hoeft}, {Horellou}, {Jarvis},
  {Kraft}, {Mevius}, {Intema}, {Murray}, {Orr{\'u}}, {Pizzo}, {Simionescu},
  {Stroe}, {van der Tol}, \& {White}}]{reinout2016}
{van Weeren}, R.~J., {Williams}, W.~L., {Hardcastle}, M.~J., {et~al.}
  2016{\natexlab{b}}, \apjs, 223, 2

\bibitem[{{Venturi} {et~al.}(2011){Venturi}, {Giacintucci}, \&
  {Dallacasa}}]{venturi2011}
{Venturi}, T., {Giacintucci}, S., \& {Dallacasa}, D. 2011, Journal of
  Astrophysics and Astronomy, 32, 501

\bibitem[{{Venturi} {et~al.}(2008){Venturi}, {Giacintucci}, {Dallacasa},
  {Cassano}, {Brunetti}, {Bardelli}, \& {Setti}}]{venturi2008}
{Venturi}, T., {Giacintucci}, S., {Dallacasa}, D., {et~al.} 2008, \aap, 484,
  327

\bibitem[{{Venturi} {et~al.}(2013){Venturi}, {Giacintucci}, {Dallacasa},
  {Cassano}, {Brunetti}, {Macario}, \& {Athreya}}]{venturi2013}
{Venturi}, T., {Giacintucci}, S., {Dallacasa}, D., {et~al.} 2013, \aap, 551,
  A24

\bibitem[{{Wilber} {et~al.}(2017){Wilber}, {Br{\"u}ggen}, {Bonafede}, {Savini},
  {Shimwell}, {van Weeren}, {Rafferty}, {Mechev}, {Intema}, {Andrade-Santos},
  {Clarke}, {Mahony}, {Morganti}, {Prandoni}, {Brunetti}, {R{\"o}ttgering},
  {Mandal}, {de Gasperin}, \& {Hoeft}}]{wilber2017}
{Wilber}, A., {Br{\"u}ggen}, M., {Bonafede}, A., {et~al.} 2017, ArXiv e-prints

\bibitem[{{Williams} {et~al.}(2016){Williams}, {van Weeren}, {R{\"o}ttgering},
  {Best}, {Dijkema}, {de Gasperin}, {Hardcastle}, {Heald}, {Prand oni},
  {Sabater}, {Shimwell}, {Tasse}, {van Bemmel}, {Br{\"u}ggen}, {Brunetti},
  {Conway}, {En{\ss}lin}, {Engels}, {Falcke}, {Ferrari}, {Haverkorn},
  {Jackson}, {Jarvis}, {Kapi{\'n}ska}, {Mahony}, {Miley}, {Morabito},
  {Morganti}, {Orr{\'u}}, {Retana-Montenegro}, {Sridhar}, {Toribio}, {White},
  {Wise}, \& {Zwart}}]{williams2016}
{Williams}, W.~L., {van Weeren}, R.~J., {R{\"o}ttgering}, H.~J.~A., {et~al.}
  2016, \mnras, 460, 2385

\bibitem[{Wright(2006)}]{Wright_2006}
Wright. 2006, The Publications of the Astronomical Society of the Pacific, 118,
  1711

\end{thebibliography}

\end{document}